\documentclass[]{emulateapj}
\usepackage{natbib,amsmath}

\begin{document}

\newcommand{\ndla}{71}
\newcommand{\kms}{km~s$^{-1}$ }
\newcommand{\cm}[1]{\, {\rm cm^{#1}}}
\newcommand{\mkms}{{\rm \; km\;s^{-1}}}
\newcommand{\delv}{\Delta v}
\newcommand{\ohi}{$\Omega_g$}
\newcommand{\lya}{Ly$\alpha$}
\newcommand{\nv}{N\,V}
\newcommand{\ovi}{O\,VI}
\newcommand{\N}[1]{{N({\rm #1})}}
\newcommand{\sci}[1]{{\rm \; \times \; 10^{#1}}}
\newcommand{\slla}{SLLS0927+5621}
\newcommand{\sllb}{SLLS0953+5230}
\newcommand{\mnhi}{N_{\rm HI}}
\newcommand{\mnciv}{N_{\rm CIV}}
\newcommand{\nhi}{$N_{\rm HI}$}
\def\fnhi{$f_{\rm{HI}} (\mnhi)$}
\def\mfnhi{f_{\rm{HI}} (\mnhi)}
\def\ltk{\left [ \,}
\def\ltp{\left ( \,}
\def\ltb{\left \{ \,}
\def\rtk{\, \right  ] }
\def\rtp{\, \right  ) }
\def\rtb{\, \right \} }
\def\nhi{$N_{\rm HI}$}
\def\lnhi{$\log N_{HI}$}
\def\omt{$\Omega_m^{Total}$}
\def\momt{\Omega_m^{Total}}

\title{Data Reduction with the MIKE Spectrometer}

\author{
Rebecca A. Bernstein\altaffilmark{1},
Scott M. Burles\altaffilmark{2},
J. Xavier Prochaska\altaffilmark{3}}

\altaffiltext{1}{Observatories of the Carnegie Institution for Science, 813 Santa Barbara Street,
  Pasadena, CA 91101, USA}
\altaffiltext{2}{Cutler Group, LP, 101 Montgomery St. \#700, SF, CA 94104}
\altaffiltext{3}{Department of Astronomy and Astrophysics, UCO/Lick
  Observatory, University of California, 1156 High Street, Santa Cruz,
  CA 95064, USA}

\begin{abstract}
This manuscript describes the design, usage, and data-reduction
pipeline developed for the Magellan Inamori Kyocera Echelle (MIKE)
spectrometer used with the Magellan telescope at the Las Campanas
Observatory.  We summarize the basic characteristics of the
instrument and discuss observational procedures recommended 
for calibrating the standard data products.  We detail the design
and implementation of an IDL based data-reduction pipeline for
MIKE data (since generalized to other echelle spectrometers, e.g. 
Keck/HIRES, VLT/UVES).
This includes novel techniques for flat-fielding, wavelength calibration,
and the extraction of echelle spectroscopy.
Sufficient detail is provided in this manuscript to enable
inexperienced observers to understand the strengths and weaknesses
of the instrument and software package and an assessment of the
related systematics.

\end{abstract}


\section{Introduction}

The field of astronomy has witnessed rapid growth over
the past few decades leading to the division of astronomers 
into observers, instrumentationalists, and theorists.  
In recent years, there is even an increased specialization 
within these sub-classes (e.g.\ numericists vs.\ semi-analytic theorists,
spectroscopy vs.\ adaptive optics imaging).
Similarly, as observatories approach billion dollar projects
the complexity of building a facility-class instruments 
exceeds the capacity of a single astronomer.
In contrast to the recent past, 
it is impractical for the majority of
observational astronomers to design and fabricate a new instrument
as the most efficient means of pursuing his/her scientific interests.  

Instead, modern observers now acquire and analyze
data-sets without having even visited the telescope or even having
developed intimate knowledge of the instrument.
Furthermore,  many observers now rely on data reduction pipelines
to produce calibrated, science-quality data without a comprehensive
knowledge of the trade-offs and limitations considered by the
software designer(s).  Although this trend toward virtual observing
may improve the efficiency with which new data is obtained
and analyzed, 
users of these tools are prone to having an incomplete
understanding of his/her experiment.

Motivated by these trends, we have written the following
paper to document the design, usage, and data reduction pipeline
of the MIKE echelle spectrometer \citep{mike}.  
We discuss observational techniques for the calibration
of these data, new algorithms to improve sky subtraction and
flat fielding, and object extraction.
Although we focus on the MIKE spectrometer, we will give 
a pedagogical discussion that generalizes to the majority of 
echelle spectrometers in use today and in the future
\citep[e.g.\ HIRES, UVES;][]{vogt94,uves}.
Portions of this manuscript also generalize to many of the low
dispersion spectrometers in use today (e.g.\ DEIMOS, FORS, IMACS).

The principal goal of this paper is to describe the data reduction pipeline
designed for the MIKE spectrometer and then generalized
to the HIRES\footnote{http://www.ucolick.org/$\sim$xavier/HIRedux}, 
UVES \citep[e.g.][]{epl07}
and ESI\footnote{http://www2.keck.hawaii.edu/inst/esi/ESIRedux/}
spectrometers within the XIDL
package\footnote{http://www.ucolick.org/$\sim$xavier/IDL/index.html}
maintained by JXP. Many of the techniques
we have implemented follow the standard lore of astronomical
research, and we have benefited from many previous
works on this topic \citep[e.g.][]{church,kelson03}.
Furthermore, our efforts have been inspired by (and take advantage of)
the algorithms developed for the spectral reductions of the 
Sloan Digital Sky Survey (SDSS; Burles \& Schlegel, in prep.).
Several unique characteristics of the MIKE spectrometer (e.g.\ tilted
sky lines),
however, have inspired new techniques for wavelength calibration,
flat fielding, and object extraction.
We describe these in detail in this manuscript; 
they may be of interest to future instruments
where similar issues arise \citep[e.g.\ X-shooter;][]{xshooter}.

The paper is organized as follows.  In \S \ref{sec:design}, we
describe the MIKE double echelle briefly and highlight points which
directly influence the design and implementation of the reduction 
algorithms described in this paper.
In  \S \ref{sec:pipe}, we describe the layout of the software pipeline.
In  \S \ref{sec:proc},  we describe the image processing algorithms.
In  \S \ref{sec:flatfield}, we describe the flat field algorithms.
In  \S \ref{sec:wave}, we describe the wavelength calibration
algorithms.
In  \S \ref{sec:object}, we describe object tracing and extraction.
Finally, we describe fluxing and coaddition algorithms in
\S~\ref{sec:endgame}.

\section{Description of the Instrument} 
\label{sec:design}

MIKE is a double echelle spectrograph which was designed for the
Magellan telescopes and installed on Magellan II (Clay) 
during November 2002.
The standard configuration for MIKE is set by the cross-over
wavelength of the dichroic 
(originally at 4550\AA\ and now 4950\AA) 
and delivers complete wavelength
coverage from 3350--5000 \AA\ (blue) and 4900--9500 \AA\ (red).  The
range can be adjusted down to $\sim3200$ \AA\ and up to $\sim10,000$
\AA\ on the red and blue sides, respectively.

In order to understand the strategies employed for observing and
for data reduction,
it is important to review the basic aspects of the optical and
mechanical design of the instrument.  We summarize  these below and
refer the reader to \cite{mike} and the MIKE user guide for
a more complete description
(http://www.lco.cl).

The first optical element in MIKE beyond the slit is the dichroic
window.  The dichroic coating on the first surface reflects blue light
and transmits red.  After the dichroic, the two arms are independent
but comprised of a similar train of optical elements.  The beam on
both sides is first converted from the F/11 focal ratio of the
telescope to a roughly F/3.5 beam by a set of ``injection'' optics
which essentially form a virtual image of the slit in the plane of each
camera, offset from the CCD.  A single set of large lenses then
comprise the collimator/camera of each arm.  These are used in double
pass to first collimate the beam on its way to the dispersion elements,
and then to focus the beam on its way back to the CCD.  The gratings are used
in quasi--littrow so that this virtual image of the slit 
and the echelle footprint are offset from each other in the focal plane of the
camera/collimator. Between the camera/collimators and the gratings,
both beams pass twice through a single prism on the red side and a
pair of prisms on the blue side for cross dispersion.



\begin{deluxetable*}{lcc} 
\tablecolumns{14} 
\tablewidth{0pc}
\tablecaption{Basic Parameters \label{tab:basics}}
\tablehead{
  \colhead{} &
  \colhead{Blue Side} & 
  \colhead{Red Side}}
\startdata                 
focal ratio (effective)& F/3.9 
                       & F/3.6            \\ 
scale at CCD           & 8.2 pix/$''$ ($0.12''/$pix)      
                       & 7.5 pix/$''$ ($0.13''/$pix)      \\
$\Delta \lambda$/pixel (unbinned)   & $\sim 0.02$\AA
                       & $\sim 0.05$\AA      \\
detector               & $2048\times4096$ (15$\mu$m pixels)  
                       & $2048\times4096$ (15$\mu$m pixels)  \\
gain                   & 0.47 e-/DN    
                       & 1.0 e-/DN        \\
read noise             & 2 e-/pix 
                       & 3.5e-/pix        \\
dark current           & 5 DN/pix/hr
                       & 2 DN/pix/hr      \\
Wavelength range       & 3200--5000\AA 
                       & 4900--10,000\AA  \\
Resolution (FWHM; $0.35''$ slit) & 83,000 
                       & 65,000           \\
Resolution (FWHM; $1.0''$ slit) & 28,000
                       & 22,000           \\
Echelle grating        & R2.4
                       & R2                \\
Prism (cross disperser)& Fused Silica, 38deg (2 prisms) 
                       & PBM2, 47deg (1 prism) \\
\enddata 
\label{tab:basic.params}
\end{deluxetable*} 


The basic characteristics of the spectrograph are summarized in Table
\ref{tab:basic.params}.  In addition to these parameters, the detailed
configuration of the dispersion elements has a variety of consequences
for the spectral format.  The most notable is that the quasi-littrow
configuration of the grating causes the slits to be tilted along the
orders so that sky lines are not aligned with the CCD rows/columns.
This is a virtue in that it guarantees that any spectral feature from
an extended source is sampled across several pixels in the spectral
direction, regardless of slit width. This improves sampling of the sky
lines and therefore sky subtraction.  Another important effect of the
configuration is that a small amount of anamorphic magnification
results in two ways: 
(1) there is a small difference in the width of the
collimated beam off the grating along individual orders, and 
(2) there is
a change in the beam width along and between orders due to angular
deviation through the prisms.  These both cause changes in the F/\#,
and therefore pixel scale.  Changes along the order are minor.
However the slit length in pixels changes by about 10\% over 
each echelle spectrogram.  
In this pipeline, we therefore quantify spatial location
along the slit in units of the slit length rather than absolute
pixel units.
This is an issue, for example, in tracing the location of the object
along the slit, as it will move due to atmospheric refraction during
any off-zenith observation.

The primary motivation for using prism cross-dispersion is the high
transmission efficiency provided.  However the angular dispersion
compared to a grating is limited by relatively small changes in glass
index with wavelength ($dn(\lambda)/d\lambda$), particularly at redder
wavelengths. This limits the order separation.  The prism apex angles
and materials were chosen to obtain a minimum order separation of
about $6''$, and a maximum slit length of $5''$ is therefore
available.  To make full use of this length, some care is taken in the
data-reduction pipeline to make 
use of the partially illuminated pixels at the ends
of the slit, which requires that the intra-order regions of the
CCD must be illuminated in the flat-fielding images.  This can be done
by inserting a holographic diffuser just after the slit (before the
dichroic) that spreads the beam coming through the slit into a 5
degree cone. This is sufficient to spread the light at any wavelength
over a 50--75 pixel region and fill the intra--order pixels.  Flat
fields taken with this slide in are referred to as ``milky flats''
or ``pixel flats'' in this pipeline.  Calibration of the
full CCD in this way also allows good measurement and subtraction
of scattered light  (see \S \ref{sec:mflat}).

The instrument is used (exclusively) in a gravity-invariant mode.  It
rests on a mounting fixture which holds it a few millimeters from the
Nasmyth mounting flange.  MIKE is not bolted to the flange because the
flange must rotate in order to move the guiders around the sky as the
field rotates with zenith angle. The field will rotate on the slit
with time as a result, with mixed implications: an off-center source
can move slightly relative to the center of the slit over a long
(20-60 min) exposure near zenith, however it is also possible to avoid
placing neighboring objects in the slit by waiting for the field to
rotate slightly.  A clear virtue of the gravity--invariant mode is
that the only mechanical motion that can occur in the spectrograph is
due to thermal variations. These thermal changes are quite minor.
The only significant motions of the spectrum on the CCD are caused by
the thermal changes of the glass and air, both of which have
significant changes in index with temperature ($dn/dT$).  The orders can
move in the spatial direction by 1--2 unbinned pixel over the course
of a night.  We track this motion by acquiring arc images in
sequence with science exposures.

The camera/collimators are mounted on rigid, flexure-mounted optical
benches.  Because the image and object of the camera/collimators are
in the same location --- the virtual image of the slit is formed by
the injection optics in the focal plane of the CCD --- one can focus
each arm by simply moving the optical bench relative to the CCD.  A
passive thermal compensation system (an invar rod with a different
coefficient of thermal expansion from the rest of the spectrograph)
keeps the instrument in good focus throughout the night, and season to
season. The same focal settings have been optimal since the
instrument was installed.

MIKE includes an internal thorium-argon lamp for wavelength
calibration and an incandescent lamp for flat fielding.  The
incandescent lamp can be used with the diffuser slide to fully
illuminate the CCD (including intra-order regions) for pixel flats, or
without to obtain `uniform' illumination of the orders that can be
used to trace their location on the detector.  We refer to the latter
as ``trace flats'', which we also use to obtain a slit profile and correct
any micro-roughness or width-variability that may exist along the
slit.  It is also possible to take trace flats using the twilight sky,
and in fact this is slightly preferable because it guarantees a truely
uniform illumination of the slit.  Although all internal lamps are
projected onto the slit with the appropriate F/11 beam (as from the
telescope), the incandescent lamp is diffused before reimaging to
avoid producing bright spots (images of the lamp filament) along the
slit.  Diffusion has been accomplished in a variety of ways over the
years and has been uniform to better than a few percent when we have
checked it.  However the twilight sky is always guaranteed to be
uniform.
It is also possible to place an iodine cell in the optical path, although
the data-reduction pipeline described here does not currently
provide routines for exploiting this option.

The pipeline includes routines for flux calibration.  As with any
spectrum taken through a narrow ($<10''$) slit, the dependence of
seeing on wavelength and atmospheric dispersion
will lead to variable slit losses.  This reduces
the accuracy of relative flux calibration over a wide range in
wavelength. However, another standard problem for echelle calibration
is variable pupil centration on the grating(s) with wavelength, field
location, and instrument alignment.  In our case, the pupil is well
centered on the grating with negligible vignetting for well-aligned
targets in both arms of the spectrograph.  However the pupil
centration can be slightly better or worse from run to run because the
instrument is not rigidly mounted to the Nasmyth rotator.  We find
that in most of our data sets taken over the first 5 years (tens of
observing runs), the centration is quite good and variability of the
flux calibration in overlapping orders is typically not worse than a
few percent.
This contrasts significantly with the results generally achieved
with other echelle spectrometers \citep[e.g.\ Keck/HIRES][]{suzuki}.

The dichroic is positioned in the optical path so that the incidence
angle of the beam is 30 degrees from normal.  The back surface of the
dichroic slide is AR coated. Unavoidably imperfect performance of both
the dichroic and AR surfaces allow for weak ghosts to be created in
the red-side spectra at the far blue range of its operation and in the
blue side.  In addition, there is the appearance of some ghosting from 
the face-on glue joints in blue side prisms.


\section{Introduction to the Data Reduction Pipeline}
\label{sec:pipe}

This section describes the layout of the code, the methods
used to organize the data, and summarizes the motivations and 
key limitations of the data reduction pipeline.
To clarify, this paper describes an IDL implementation of
a MIKE reduction pipeline.  There is an entirely separate pipeline
based on python that was written by D. Kelson 
(http://www.lco.cl/telescopes-information/magellan/instruments/mike).
Throughout the algorithms, the detector is oriented in memory 
within IDL such that rows correspond to spatial information and columns
correspond to spectral information.

\subsection{Setups}

The previous section discussed the general design of the
MIKE spectrometer.  Common to most echelle spectrographs,
there are only a few optical elements or detector configurations
that can be adjusted by the observer.  For MIKE, these are the
CCD binning, slit plate position to choose the illuminated slit, 
and the angles of the dispersing elements.
In practice, the dispersers are rarely adjusted
because the standard setup provides nearly continuous wavelength
coverage from $\lambda \approx 3250-9000$\AA.  

Throughout the remainder of the paper, the term ``setup'' defines
a unique configuration of the slit, CCD binning,
and disperser angles.  To properly reduce the data, one requires
a full set of calibration exposures for each setup.  
In this sense, one must treat the observations and data reduction of
each unique setup separately. 
Because the slit is adjusted through an interface separate from the CCD
controller interface, the slit plate position is not
recorded in the FITS header.  Although the slit could be
determined automatically (e.g.\ by measuring the FWHM of the arc lines),
we require the user to designate 
which exposures correspond to which setup.

\subsection{Pipeline Organization}
 
Basic characteristics of each data frame (i.e.\ each FITS file)
are parsed from the FITS header and recorded in an 
IDL structure (akin to a structure in C)
in a series of ``tags''.  These include the right ascension (RA),
total exposure time (EXP), CCD binning 
(COLBIN, ROWBIN), etc.  This structure is archived as a binary FITS table
that one can edit with standard FITS tools (e.g.\ {\it fv})
or simple IDL scripts.  The observer is required to identify the
science and calibration frames of a unique setup (via the SETUP tag)
and must identify and flag any corrupted, junk, or test frames.

The pipeline structure is called into memory within the IDL package
and is passed to nearly every algorithm. 
This includes an algorithm which automatically identifies the 
exposure type (e.g. {\it BIAS, FLAT, ARC, OBJ}) according to the
length of the exposure, the total counts, and
level of structure in the image.
Science exposures of a given SETUP
are further parsed by the object name (from the FITS header) and 
each unique target is assigned an integer OBJ\_ID value. 
Most of the algorithms are then controlled with the
IDL structure, the SETUP number, and the OBJ\_ID value.
All of the tasks can be carried out on the two cameras
independently, as desired.

\subsection{Motivations and Limitations}

The design of the reduction pipeline was guided in large
part by our scientific interests.  Our primary targets are
relatively faint ($V>17$), extragalactic sources with negligible
(quasars) or small spatial extent (globular clusters).
Our science goals generally required modest S/N ($\approx 10-30$)
and moderate spectra resolution ($R\approx 30,000$). 
None of our science goals required accurate absolute fluxes.
These characteristics describe many of the programs carried
out with MIKE to date \citep[e.g.][]{cpb+05,mlk+07,dab+07,fpl+08},
but there is also a substantial community interested in very
high S/N and/or high precision observations.

With the above considerations in mind, we set out to achieve
the following goals with our observing and data reduction
procedures:

\begin{itemize}

\item  Achieve a relative wavelength error of less than 0.05\,pix
and an absolute wavelength scale of better than 0.2\,pix.

\item  Achieve better than 10$\%$ precision in the relative fluxing
of objects.  This is primarily  for co-adding 
overlapping echelle orders.

\item  Implement bias subtraction and flat fielding procedures
that remove pixel-to-pixel variations to better than 1$\%$ statistical
error without introducing $1\%$ systematic error.

\item  Maximize the number of pixels analyzed along the spatial 
dimension for sky subtraction.  This is especially important for MIKE
because of its relatively short slit length ($5''$).

\item  Approach the Poisson limit with sky subtraction and extraction.

\item  Perform an accurate measurement of the spatial profile and
its variation with wavelength.

\item  Derive a 1D variance array which describes the principal sources
of error (Poisson noise, read noise, flat fielding).

\item  Robustly flux calibrate and combine multiple exposures across
diffraction orders to create a single 1-dimensional flux calibrated 
spectrum for each object.

\end{itemize}
Of these, the most difficult requirement is flat fielding:
There are challenges to obtaining appropriate calibration data 
(e.g.\ achieving sufficient counts at 3200\AA) and also
detector blemishes (especially on the blue side) 
that introduce $\approx 1\%$ systematic errors.

Granted the relatively narrow scientific programs which
motivated our observing strategy and pipeline design,
we caution about the following limitations.
First, we have not carefully explored or addressed 
systematic errors at levels of $\approx 0.5\%$.
For example, detector blemishes (pock marks) are present in 
the red CCD.  These are easily identified yet difficult to
calibrate to better than the $1\%$ level.  Furthermore,
we have not demonstrated that the algorithms which subtract
scattered-light perform significantly better than $1\%$ precision.
Therefore, it may be difficult to recover very high S/N data
($\gg 100$) with our procedures.
Second, we have not developed algorithms to properly subtract
the sky background for observations of very bright
or spatially extended objects (i.e.\ where the object 
exceeds the sky flux in every pixel along the slit).
We have assumed that at least a few pixels in the slit are
sky dominated or we skip sky subtraction altogether.
Third, we have not introduced algorithms to calibrate
the iodine cell. 

\section{Image Processing}
\label{sec:proc}

The first step in all of the CCD pipeline routines is standard
digital processing.  In a custom routine, called `mike\_proc',
we process both red and blue CCD data. 
This routine includes the following steps applied in sequence:
(1) remove smooth overscan levels in both columns and rows;
(2) trim the data to the active 2048x4096 native pixels;
(3) apply the pixel-to-pixel flat; 
(4) convert ADU to electrons by applying the gain; 
(5) calculate and record the inverse variance of the image, 
including readout noise. 
The final step adopts an algorithm which 
approximates the noise in the low count (i.e.\ Poisson) limit.

Each raw CCD frame is read into memory as standard unsigned 16-bit integers 
in the same state as it was saved at the telescope.  The overscan regions
and on-CCD binning is deduced directly from the raw image size
under the assumption that the full frame was read to disk.
For standard MIKE pipeline reductions, we do not subtract bias frames nor
dark frames.  We have constructed super bias and super dark frames, but no
features were apparent and not nearly as significant as the structure that
appears and varies in a frame-by-frame basis.  

\begin{figure*}
\includegraphics[height=7in,angle=90]{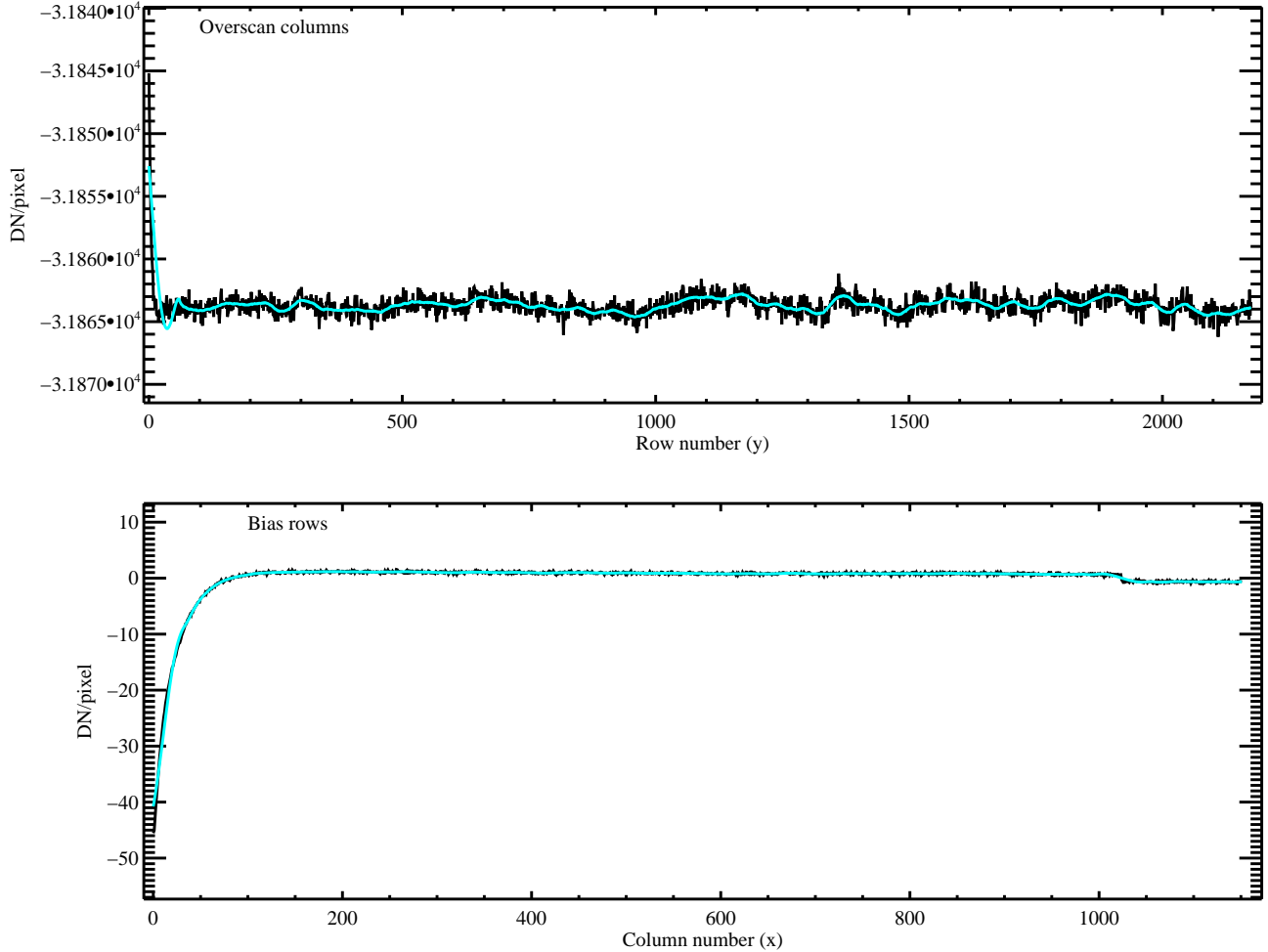}
\caption{Upper panel: Sigma-clipped average of the overscan columns
for a red-side frame (black data).  Overplotted on this overscan
average is the same data filtered by a SAVGOL function with 30 pixels
(121 total native pixels)
on each side.  This method allows us to produce a smooth function
that also recovers sharp features. 
Lower panel: Bias row from the same CCD.  Overplotted is the 
data smoothed with a symmetric 37 pixel (73 native) SAVGOL filter.
}
\label{fig:bias}
\end{figure*}

We need to carefully model the overscan regions
to remove the digital features in the CCD image.  The first overscan
region that is modeled includes all columns that readout 
immediately after each CCD row.  First, we 
average these columns with 3-sigma clipping to produce a mean 1D 
overscan spectrum as a function of CCD row.
Then the routine convolves this spectrum with 
a 4th degree Savitzky-Golay filter having 
a symmetric width of 121 native pixels average 
(Figure~\ref{fig:bias}; upper panel). 
This averaged, filtered spectrum is replicated and subtracted from
the entire, raw CCD frame.  
The same procedure is then applied to the set of rows 
immediately preceding the active CCD rows.  Similarly, a 73 native pixel 
Savitzky-Golay filter is convolved with the average overscan in each column
(Figure~\ref{fig:bias}; lower panel).
This result is replicated and subtracted from the entire CCD frame.

There have been two CCD detectors used with the blue side of
the MIKE spectrometer.  
At commission, a SITe ST-002A CCD with relatively poor quantum
efficiency (QE) at $\lambda < 3800$\AA\ was installed.  This CCD was
replaced on May 6, 2004 with a Lincoln Labs CCID-20 device which has
high QE down to the atmospheric cutoff.  However, all data taken with
the upgraded blue CCD between the dates of May 6, 2004 and September
21, 2005 suffer from a non-linearity in count rates.  On the latter
date, the voltages were reset to correct this non-linearity.  To deal
with this non-linearity in affected data, we adjust the raw CCD counts
after overscan subtraction by multiplying this correction factor (CF)
to the processed electron counts on a pixel by pixel basis:
\begin{eqnarray}
{\rm CF(Image)} &= 1, \; {\rm if \; Image} < 14,000 \;{\rm ADU} \\
                        &= 2, \; {\rm if \; Image} > 35,000 \;{\rm ADU} \nonumber\\
                        &= \exp{\left(\frac{{\rm Image}-14,000}{25,000}\right)^2}, 
                            \;\;\; {\rm otherwise.}\;\;\; \nonumber
\end{eqnarray}
Any image pixel corrected to values over 50,000 electrons is capped 
artificially at that value, as
this non-linearity correction is not reliable for such large corrections
and one can assume these pixels are saturated.

After overscan subtraction, the image is trimmed to active CCD pixels
and zero now represents the average level of the overscan regions.  
Assuming a flat-field of the pixel-to-pixel response has been created
(see \S\ref{sec:mflat}), the image is normalized directly by the pixel flat.
The CCD gain is measured by assuming Gaussian statistics in pairs of flat-field
images, and is applied (\S\ref{sec:mflat}) 
so that all CCD images are converted from digital counts (ADUs)
to electrons measured.
We estimate the error on each pixel and report this as the inverse
variance: $1/\sigma^2$.  
The first estimate of this error is an approximation
based solely on the measured number of electrons in each pixel as well as
the global estimate of the readout noise in electrons.  Gaussian statistics
dictates that the variance is simply the number of counts in the expectation.
For faint sources, the data provide only a very
noisy estimate of the count rate. 
More importantly, one encounters a large
systematic bias in the estimated error as the counts approach zero.  
In addition, the shape of the likelihood function starts diverging 
significantly from a simple Gaussian which is assumed in the rest of 
the pipeline.  In order to ameliorate these effects we approximate the
variance with this function:
\begin{equation}
\sigma^2 = \left|{\rm Image(e^{-})} - \sqrt{2}\,RN\right| + RN^2,
\end{equation}
where Image(e$^-$) is the observed number of electrons and
RN is the image readout noise measured in electrons.
At very small Image(e$^-$) values, this approximation avoids
having unreasonably small variance values.
The final correction is simply an addition in quadrature with
the variance added by using a pixel flat.
This correction can be minimized by taking a long series of flat exposures.

The last step in the processing is to identify bad rows, passed in with 
the BADROWS keyword and subsequently set the inverse variance to 0.
For MIKE, the first two and last two rows of each CCD seem to always
carry excess counts and are flagged with zero inverse variance (zero weight)
by default.

\section{Flat-field Calibration}
\label{sec:flatfield}

In this section, we describe the flat-field calibration frames
and the algorithms which process and analyze these data.  We
use the frames for a variety of purposes, including 
(1) classic ``pixel flats" for image correction of relative 
pixel-to-pixel CCD response, 
(2) ``trace flats" for definition of the spatial echelle order boundaries, and
(3) ``slit flats" for the characterization of the spatial profile 
of a uniformly illuminated slit in each echelle order.
We describe the implementation of each of these steps in detail below.

\begin{figure}
\includegraphics[width=3.5in]{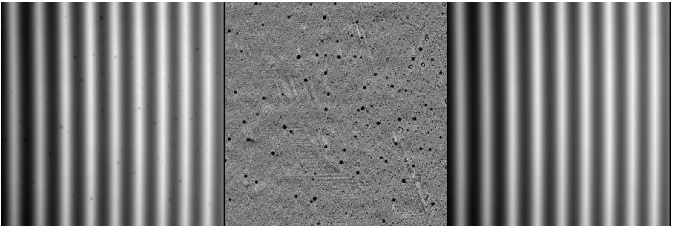}
\caption{Milky flat processing: The left panel shows a combined milky-flat
image subsection cleaned of cosmic rays (white is higher intensity).  The
middle panel shows the pixel-to-pixel flat field, normalized to unity.  The
right panel is the left panel frame corrected by the pixel-to-pixel flat.
This is identical to the smooth model constructed from the milky-flat. }
\label{fig:mflat}
\end{figure}

\subsection{Pixel-to-Pixel flats}
\label{sec:mflat}

The first flat-field calibration routine, mike\_mkmflat,
produces a relative pixel flat-field image for both the blue 
and red cameras.  The routine is straightforward: We use a stack of
MIKE images obtained with the `milky diffuser' in the beam directly behind
the slit to obtain smooth flux distributions in both the spatial and 
spectral directions (these frames are referred to as milky flats).
The practice of acquiring and analyzing a milky flat for pixel-to-pixel
variations contrasts with standard practice in echelle spectroscopy.
It has several advantages.  First, we can correct for pixel sensitivity
variations at and beyond the slit edge.  This allows us to include
several pixels at the slit edge boundary during sky subtraction
and extraction.  Second, stray light (including ghosts) which often exhibits
sharp features at random angles with respect to the echelle orders
is smoothed out and its effects are minimized.  
Third, it is easier to calibrate larger defects in the CCD and
especially defects which partially cover an echelle order.
Finally, one is assessing the pixel-to-pixel quantum efficiency with
the same color light as the science exposures.
Ideally, a stellar source is used in the blue milky flat to maximize
counts at 3300\AA\
and the internal flat-field lamp is used in the red  
to minimize the effects of sharp features (e.g.\ from atmospheric absorption
features).

Each milky flat image is processed as described in $\S$~\ref{sec:proc}
and the stack of images is co-added to produce a single milky flat image 
for each CCD 
(a separate set of milky flats is required for each 
CCD binning used during data acquisition).
We remove the large-scale flux variations as a function of position on the
CCD (the routine default is to resample at the 256 native pixel scale).
Then we apply a standard median filter in 1-dimension along CCD columns
(in the spectral direction with a default width of 45 native pixels),
on the assumption the calibration source has no small scale spectral features.
This assumption breaks down for astrophysical sources with 
strong atmospheric telluric absorption bands observed with the red CCD.
A 2-dimensional median filter (default median box is 9 native pixels on a side)
is applied to the normalized milky-flat image processed above,
and serves as a proxy for a model with which to reject the strongest features 
in the flat-field.  Normalized milky-flat pixels are masked if they or a 
neighbor inside the median box size deviate by more than 3\%.  Finally, these masked
pixels are replaced with a smoothed version of the normalized flat (with masked pixels given zero weight, and a 2-d smoothing length of twice the median box).  
The processed milky-flat is normalized by this final smooth image to 
effectively remove the remaining large-scale fluctuations and the resulting 
image is stored as the normalized pixel-to-pixel flat.  
In Figure~\ref{fig:mflat}, we 
show an example of the normalized flat-field of the red CCD in the middle panel.
The co-added milky-flat image is shown before (left panel) and after 
(right panel) applying the processed pixel flat.  
One will notice that there are a large number of ``pock" marks 
(moderate flat-field features) on the red Site CCD and 
the original blue Site CCD.  They cover about 40 native pixels each and have an 
areal density of 5000 pixels per CCD.    
Reconstructing the pixel-to-pixel response as shown in Figure~\ref{fig:mflat} 
is limited by a systematic error of approximately 1\% per pixel.
The statistical uncertainty in the pixel flat is saved, and applied to
the statistical error (the inverse variance) on all images when 
the pixel flat is applied.

We calculate the gain between pairs of milky flats, with the estimator
provided below.  It is calculated from each pair of milky flats in 
sequence and the mean value is recorded for each image taken with this setup.
A comparison image is made from the two flat images, $i1$ and $i2$, as such:

\begin{eqnarray}
image = & \sqrt{\frac{1}{\frac{1}{i1} + \frac{1}{i2}}} \, 
              {\rm ln}[\frac{i1}{i2} \, \bar{\left(\frac{i2}{i1}\right)}]  \cr
1 / <gain> = & variance(image) .
\end{eqnarray}
For a series of $N$ milky flat images, $N-1$ sequential pairings are analyzed,
and the mean gain calculated from the $N-1$ pairings is used as the gain 
for all of the images being processed (with appropriate binning and side).

\subsection{Trace Flats}
\label{sec:tflat}

A short set of direct flat images (without the diffuser behind the slit) is
used to both define the echelle order boundaries and the relative spatial
cross-section of the slit profile.  The source can either be a short
exposure of the internal flat-field lamp or the twilight sky.  
The internal lamp is better suited for defining the order boundaries, and
the twilight sky provides a better direct illumination of the focal plane.

\begin{figure}
\includegraphics[width=3.5in]{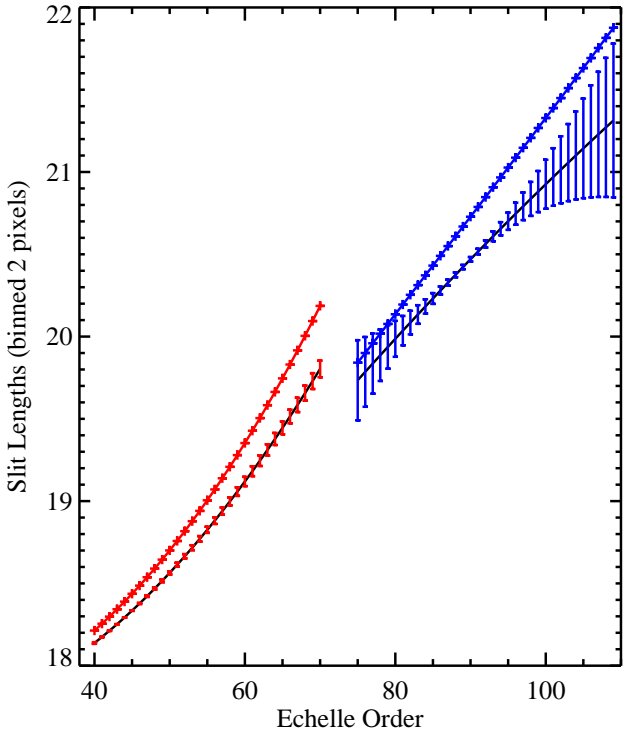}
\caption{Length of the standard 5$''$ slit in pixels (binned 2x from the
native $15\mu$m pixels) as a function of echelle order.  The lower curves do
not take into account the tilt of the slit, i.e.\ it represents the slit
length projected onto the rows of the CCD.  The upper curves account
for the tilt of the slit in the CCD frame. 
The bars indicate the variations within a given order.
}
\label{fig:slitlen}
\end{figure}

\begin{figure}
\includegraphics[width=3.5in]{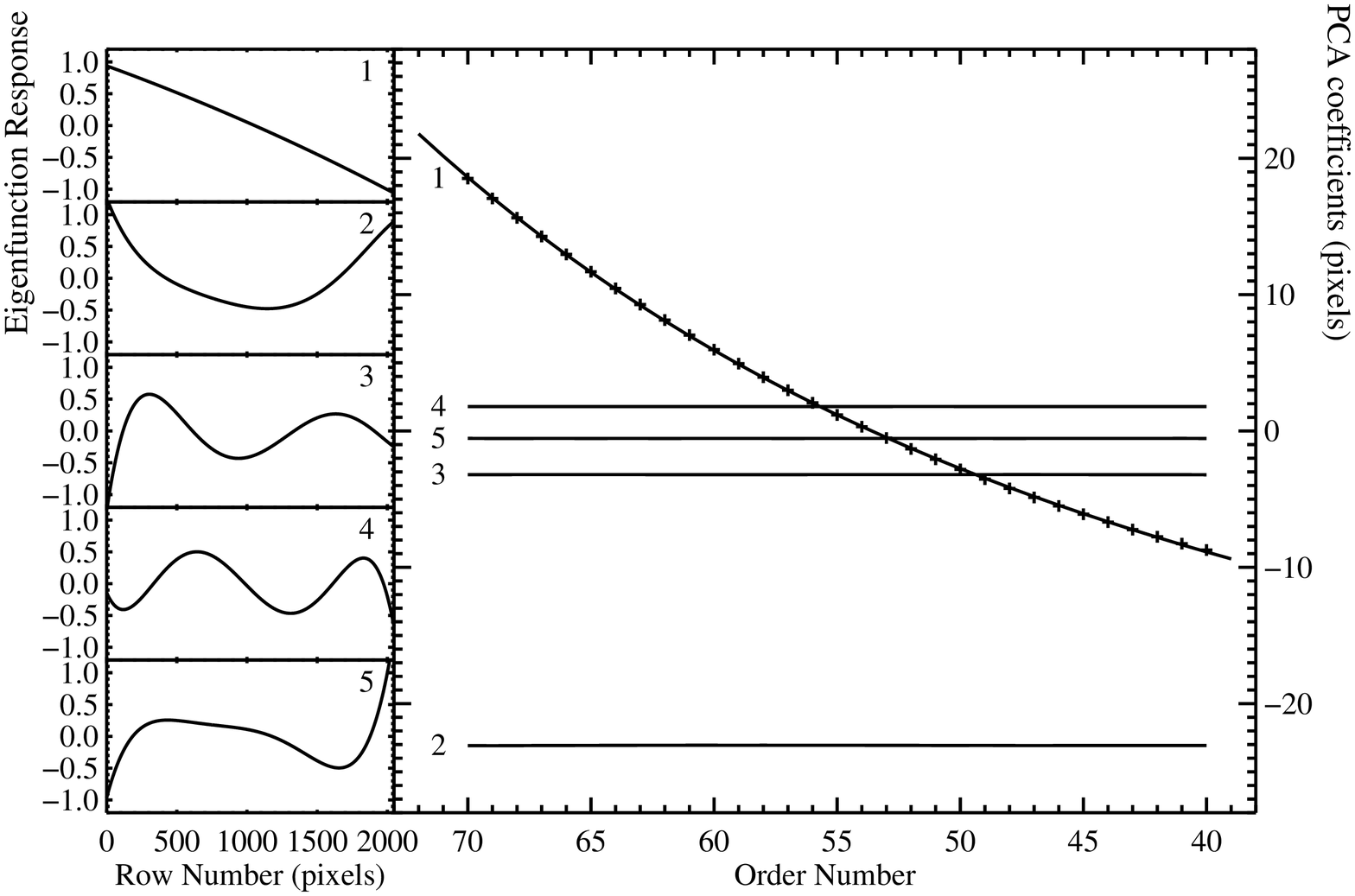}
\caption{Principal component analysis of the Legendre polynomial coefficients
that describe the order centers of each echelle order as a function of CCD row
from a representative red-side frame.  The left-hand panels show the
PCA eigenfunctions, normalized from $-1$ to 1.  The right-hand panel
shows the PCA coefficients as a function of echelle order.  The code
fits a 2nd order polynomial to the first PCA coefficient in order to
interpolate bad echelle orders or extrapolate to overlapping orders or
regions with faint signal.  One notes that the coefficients are essentially
constant for PCA coefficients 2-5.
}
\label{fig:pca}
\end{figure}

\begin{figure}
\includegraphics[width=3.3in]{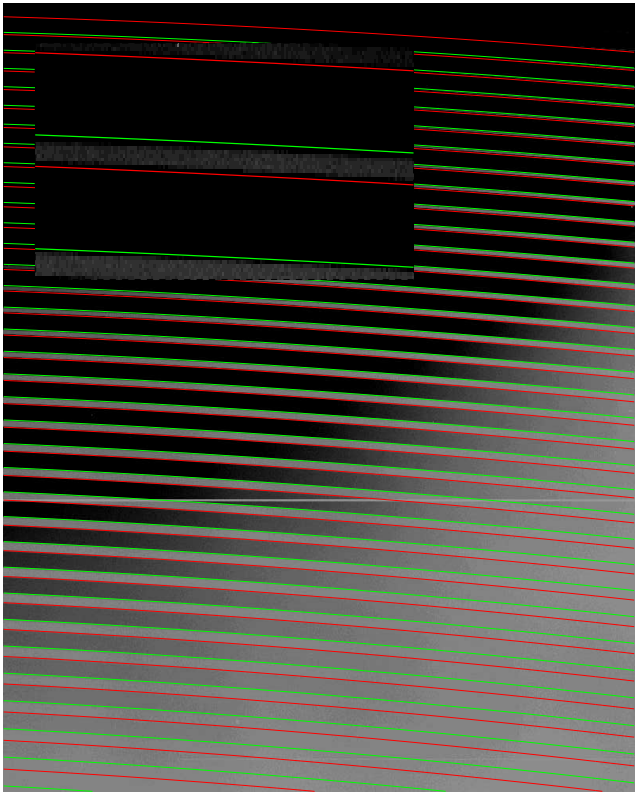}
\caption{Image of a trace flat (black is high intensity) from the blue-side
of MIKE.  The bluest echelle orders are to the bottom.  The inset shows
a zoom-in of a portion of two echelle orders.  The green and red lines
trace the bottom and top sides of each echelle order, respectively.  Note that
the order edges have been extrapolated into the bluest regions of the
CCD (lower right in the figure).  
The PCA analysis provides a robust means for performing
this extrapolation.}
\label{fig:trace}
\end{figure}

As with the Milky flats images above, a series of direct flat images
are combined to yield a cleaned red and blue
direct flat image, which we refer to as the combined ``trace flats".
The peaks in each trace flat image are located after applying a sawtooth filter
in the spatial direction (along rows).  This procedure effectively locates
the 50\% level of both the left (positive peaks) and right (negative peaks)
edges of the echelle orders in the trace flat images.  Echelle orders in which
90\% or more of the order was traced along the length of the CCD are considered good, and used to predict the edges of the remaining order.  
We then calculate the center and width of each echelle order, which are
functions of CCD row number.
The order width varies very slowly across the
CCD array (see Figure~\ref{fig:slitlen}), and can be described by a 2-dimensional
polynomial which is linear with respect to row number and quadratic with 
respect to order number.  
We fit the order center with a Legendre polynomial as a function
of CCD row number. We then use standard
PCA analysis \citep{PCA} of these order centers 
by searching for the principal components in the 
polynomial coefficients that describe the central column positions of the 
good orders.  
An example of these normalized basis functions are shown in Figure~\ref{fig:pca} 
in the 
left-hand panels (above and beyond the 0th order shape which is simply the 
mean order position on the CCD).  The linear combination of the functions when
multiplied by the smoothly varying coefficients shown in the main panel of
Figure~\ref{fig:pca}
accurately describe the order centers (for all rows and all orders) on the CCD.
To predict the locations of the remaining orders which fall on the CCD, 
but which had poorly defined edges, we simply use the extrapolated
coefficients based on the fits to the well defined order centers.
Figure~\ref{fig:trace} shows an example of the order definition for
a trace flat observed with the blue camera.

One might consider a slightly more simple procedure which is to 
extrapolate the polynomial coefficients describing the order centers 
as a function of order number.  
This usually produces unsatisfactory results due to the highly
correlated condition of the coefficients.  The principal components are
defined to be orthogonal by definition, and the variation of each 
coefficients can be reliably extrapolated independently of the others.
In the upper right corner of Figure~\ref{fig:trace}, one notes 
order traces extrapolated into regions with nearly negligible flux.
The decomposition of the left and right edges into order centers and
order widths produced very reliable edge definitions, and 
allows for automated localization of the 
orders with no human interaction.

The pipeline routine responsible for determining the edge positions
is called, mike\_edgeflat, which produces an order structure with the
prefix:  OStr\_ in the Flats directory.

\subsection{Slit Profile Correction}
\label{sec:slitprof}

The most involved correction derived from the
flat-field calibrations is the slit profile correction, which we refer
to as the ``slit flat".  Once the echelle orders have been defined, and the
wavelength image has been created from arc frame images 
(see $\S$~\ref{sec:arcimg})
we can measure the relative throughput of the spectrograph as a function of
position along the 5$''$ slit length.  
Although simple in principle, we invoke a number
of steps in order to reliably measure the slit profile.  In the case of no
correction, a spatially uniform source (like the dark night sky), would
exhibit no change in the measured flux as a function of slit position.
In order to correctly model the slit profile, the counts must approach zero
in the gaps between the orders.  A smooth scattered light image is constructed 
by fitting to the empirical counts in CCD pixels not associated with any
echelle orders.  The routine (x\_modelslit)
fits the scattered light row-by-row with 
8 Legendre polynomial coefficients. 

Each pixel in the trace flat is assigned three values in addition to the counts
measured in the combined direct flat:  order number, relative wavelength
(recorded in pixels), and spatial slit position.  
The first two numbers are derived by the wavelength calibration
procedures in the next section.
The spatial slit position is measured as a quantity
called slit fraction, which is normalized between -1 and +1, 
where -1 represents a pixel position exactly at the left edge 
of the slit-length, 
+1 is the right edge, and 0 is a pixel at the exact center of the slit.
More precisely, these values are in units of half slit-length (HSL).
A relative wavelength is given to each pixel by correcting for the tilt
of the slit in the spectral direction and distance from the order center.

\begin{figure}
\includegraphics[width=3.3in]{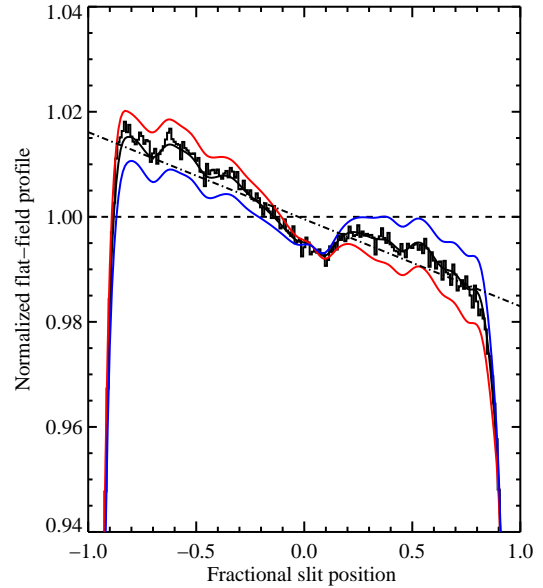}
\caption{Slit profile determination for the 67th echelle order on the red side
of MIKE.  The three profiles (red, black, red)
represent the profile fit for the top, middle and bottom sections of echelle order 67,
with the smooth solid lines the b-spline function evaluations for each case.
}
\label{fig:slitflat}
\end{figure}

Order by order, the counts in each pixel are normalized by 
fitting a b-spline set of 
coefficients as a function of relative wavelength to the pixel counts
in the central 70\% of the slit ($-0.7 < HSL < +0.7$).
The b-spline breakpoints are 
separated by 1.2 CCD pixels in the spectral direction, which is sparse enough
to suppress the high-frequency noise but dense enough to fit real 
spectral features in the twilight sky.
The smooth b-spline spectrum is evaluated at all pixels falling in the current
order, and the counts in each pixel are normalized by this evaluated fit.
In addition, the inverse variance in each pixel is scaled accordingly to 
preserve the reported signal-to-noise in each pixel (Figure~\ref{fig:slitflat}).

Now the spatial profile in each echelle order is fit in two steps.  The first
step is to fit the mean profile, which is accomplished by fitting a single
b-spline function to all of the normalized counts in the order as a function of 
slit fraction.  The b-spline is evaluated at unit intervals of 0.01\,HSL and stored
from $-1.25$ to +1.25\,HSL.  The 251 samplings of the mean profile per order
is stored in the structure with the tag PROFILE0.
Any trend as a function of wavelength is fit with a linear dependence of
row number and a b-spline fit to the residuals of the normalized trace flat
about the mean evaluated b-spline.  The typical linear deviations are 
around the 10\% level along the length of a single order.
These linear coefficients are sampled at the same 251 slit positions and
stored as PROFILE1.  One can reconstruct the appropriate slit profile
for all rows in the order, for instance, the bottom row is given by 
PROFILE0 - PROFILE1 and the top row by PROFILE0 + PROFILE1.

The slit profile in each echelle order 
and the gradient of the profile with respect to position along the order
are used to correct the spatial flux profile of the night sky before 
object extraction.
The slit profile is not applied directly, given the low S/N region beyond 
the order boundaries, but rather the profile is used in the model basis 
of the sky background.  We do find a difference in the recovered slit profile
when a direct flat of the internal calibration lamp is compared
to results of a direct flat of the twilight night sky.  There is a significant
linear gradient across the extent of the slit in the internal lamp images
because of non-uniform illumination, and
this slope is removed in the stored profile.

\section{Wavelength Calibration}
\label{sec:wave}

This section describes the series of algorithms that calibrate the
wavelengths for every pixel that falls within the
echelle footprint.  Following standard practice, the wavelength
calibration is derived from the spectrum of a ThAr lamp observed with
the same setup as the science exposures.  While the
spectrograph in standard observing mode does not physically move, and so
does not have flexure, the spectrum can move on the CCD by several
native pixels over a night due to changes in the temperature of the
air, glass, and metal in the spectrograph.  Coeval calibrations are
therefore important.  Below, we will describe both the calibration of
individual ThAr spectral images (``arcs'') and the strategy we employ for obtaining
shifts in the echelle footprint over a night, and between 
science exposures and their ``coeval'' ThAr exposures.

Our approach to wavelength calibration differs from standard practice
in that we generate a 2D wavelength image from each arc frame.  To do
so, we begin by deriving a 1D wavelength solution along the spatial
center of each echelle order.  We then trace the tilt of the ThAr lines
out to the spatial edge of the order, and finally propagate the 2D
solution for all pixels that fall within an order.  It is worth noting
that the tilt in the wavelength solution is advantageous to the data
analysis in that it improves the sampling of sky lines, particularly
near the Nyquist limit (i.e.\ narrow slit).  Our specific treatment
of this tilt is necessary because the tilt varies along and between
echelle orders, and must be measured and accommodated across the full 2D
spectrum to accurately subtract  sky lines and 
extract the data.

\subsection{Thorium-Argon Calibration Data}

An internal Thorium-Argon (ThAr) hollow cathode lamp is used to 
obtain the required intensity and
line density as a function of wavelength for calibration. When an arc
image is taken, the switch that turns on the lamp also moves a small
mirror into optical path between the telescope and the slit plate,
blocking the light from the sky and allowing an F/11 image from the
lamp (a bright spot, roughly 1 cm in diameter) to be projected onto the
slit instead. This configuration is difficult to miss in the picture
displayed by the slit-viewing camera and, as such, acts 
as a convenient safety feature for
distracted observers who might otherwise take an arc simultaneously
with a science exposure.

An exposure of 1-5 seconds provides sufficient lines for calibration
of the full spectral range for all slit widths (0.35--2 arcsec).  Our
standard calibration template includes 12 lines in the bluest orders
and 20--30 lines per order over most of the optical range.  Strongly
saturated lines dominate the appearance of the arcs around
7500\AA. These do bleed electrons into neighboring columns, but
they do not pose a calibration problem.  A bigger issue is that the
useful lines for calibration in the reddest orders 
($\lambda \gtrsim 7800$, 
red order number less than 44) are fairly weak.  Exposure times on the
red side should be dictated by these weak lines if the reddest orders
are of scientific interest.

In this pipeline, all wavelength calibrated data is converted from
in-air wavelengths to vacuum wavelengths by default.  It is important
to realize that the change in the photon wavelength when traveling
through a medium is a function of wavelength, unlike a Doppler shift,
so that this conversion will affect any analysis, even if a Doppler
shift is allowed.\footnote{The relationship between the wavelength of
  light in vacuum, $\lambda_{\rm vac}$, and in a medium of index, $n$,
  is $\lambda = \lambda_{\rm vac}/n(\lambda)$.  For air, $n-1 =
  6.4328\times 10^{-5} + 2.94981\times 10^{-2}/(146-\sigma^2) +
  2.5540\times 10^{-4}/(41-\sigma^2)$, 
  in which $\sigma=10^4/\lambda$ \citep{ciddor96}.}
This conversion can be avoided, as desired, 
but the reason for making vacuum
wavelengths the default is that while optical--wavelength lines are
typically discussed at ``in-air'' wavelengths, UV--wavelength lines
are always discussed at ``in--vacuum'' wavelengths. 
Consequently the
standard for galaxy and stellar spectroscopy is in--air wavelengths,
while the standard for QSO absorption line observations is in-vacuum.
To understand the implications of this conversion, it is important to
think about how the lines are produced and how the data are
recorded.  First, the photons from both astronomical sources and
calibration lamps are produced in a vacuum, travel in air through the
spectrograph, and are detected in a dewar under vacuum\footnote{As a
  pedagogical point, we note that the only record of the wavelength of
  a photon in a spectrograph is its {\it location} on the CCD, which
  is determined by the angle of the photons after reflection from the
  grating.  Thus, any spectral features from any source will be
  recorded at in--air wavelengths.}.  The wavelengths designated to the
lines in the arc spectrum will determine the definition of the
wavelength calibration, and it will be appropriate for both
astronomical sources and calibration lamps.  If the line list for the
arc lines gives in--air values, then the calibration is in-air and
must be converted to vacuum to discuss UV wavelength lines.  

\begin{figure*}
\includegraphics[height=7.0in,angle=90]{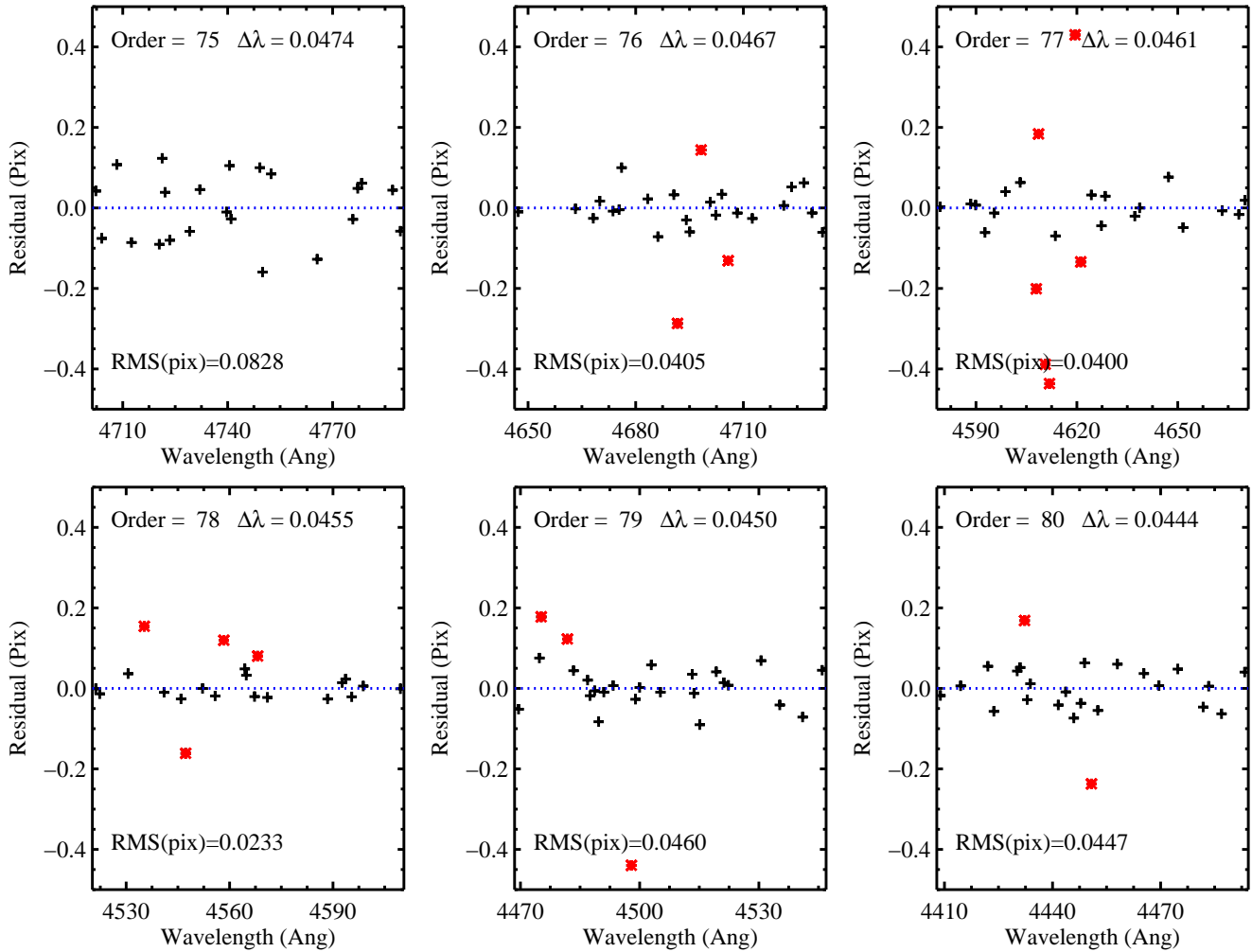}
\caption{Residuals (in pixels) from Legendre fits of auto-identified
and centroided ThAr lines in a series of echelle orders from the
blue camera of MIKE.  The red 'stars' indicate lines rejected by
the algorithm during fitting.  In general, we used 4th order
polynomials.  This is an example of the quality assurance (QA) 
files produced by the pipeline.
}
\label{fig:wav1D}
\end{figure*}

\subsection{Standard Calibration}

The first algorithm follows standard practice for echelle reduction.
We perform basic overscan and bias subtraction and then flatten the
blue CCD as described in the previous sections.  With the red CCD,
however, we skip the overscan subtraction (although the average value
is written to the bias row) because the very bright arc lines near
above 7500\AA\ tend to bleed into the overscan region.  The code then
calculates the image shift between the arc frame and the trace flat
solution as described below.

For every echelle order, the code performs a pseudo-boxcar
extraction by rectifying the order with linear interpolation
and extracting the central column of each order.  
The individual order extractions are stitched together to 
create a single array.  This array
is cross-correlated against the same construction from 
an archived, calibrated image using an FFT algorithm.
The peak in the FFT sets the shift between the stitched 1D
arcs of the archived and new arc images. 
The derived shift is then used to identify the physical order
numbers of the new arc 
and to also predict the average shift in the spectral direction
(e.g.\ due to changes in the echelle angle). 

Each echelle order is then analyzed individually.  Its 1D spectrum
is cross-correlated with an FFT to the single-order 1D archived
spectrum to improve the precision of the spectral shift.
The code then derives a wavelength solution for the new 1D spectrum
assuming the archived solution with the derived offset.
The algorithm then identifies all ThAr arc lines using a peak
finder which demands a high signal to noise ratio (SNR) 
and a `pointed' feature whose
fluxes in the central five pixels $f_i$ are appropriately rank ordered
($f_1 < f_2 < f_3 > f_4 > f_5$).
Those lines that satisfy this rank ordering
and lie within three binned pixels of a laboratory-calibrated
ThAr line are centroided and recorded.
The line centroid is measured by (i) splining the flux of the
central seven pixels, (ii) evaluating the peak of the spline,
(iii) identifying the left-hand and right-hand edges of 
the spline corresponding to 33\% of the peak, and (iv) averaging
the two edges to derive the centroid.
This algorithm is non-parametric and robust.
The result is a
high SNR subset of ThAr lines with measured pixel centroids
and known laboratory wavelengths.

The code then performs a low order, Legendre polynomial
fit with aggressive rejection
to the pixel values versus the logarithm of the laboratory
wavelengths.  By imposing the logarithm of the wavelength, 
one significantly reduces a systematic bias 
which favors long--wavelength lines.
This is less important for individual echelle 
orders (where the wavelength range is only
$\approx 100$\AA), but it is critical for the 2D polynomial
fit described below.
Finally, the code again searches for ThAr lines in the 1D
spectrum but using this fit for the wavelength solution and also
allowing lines with lower SNR in the image.  A final 4th or 5th
order Legendre polynomial fit is derived, again with aggressive
clipping.  The final output for each order
includes the 1D spectrum and the pixel centroids and ThAr wavelengths
of all the `good' ThAr lines that were not rejected during
the Legendre polynomial fits.
Figure~\ref{fig:wav1D} presents the residuals for Legendre
fits to echelle orders 75 to 80 from the blue camera of MIKE.
In nearly every individual order, we achieved line residuals 
of RMS$<0.1$\,pix
and a fit solution with RMS$<0.05$\,pix independent of binning.

\begin{figure}
\epsscale{1.2}
\includegraphics[height=3.5in,angle=90]{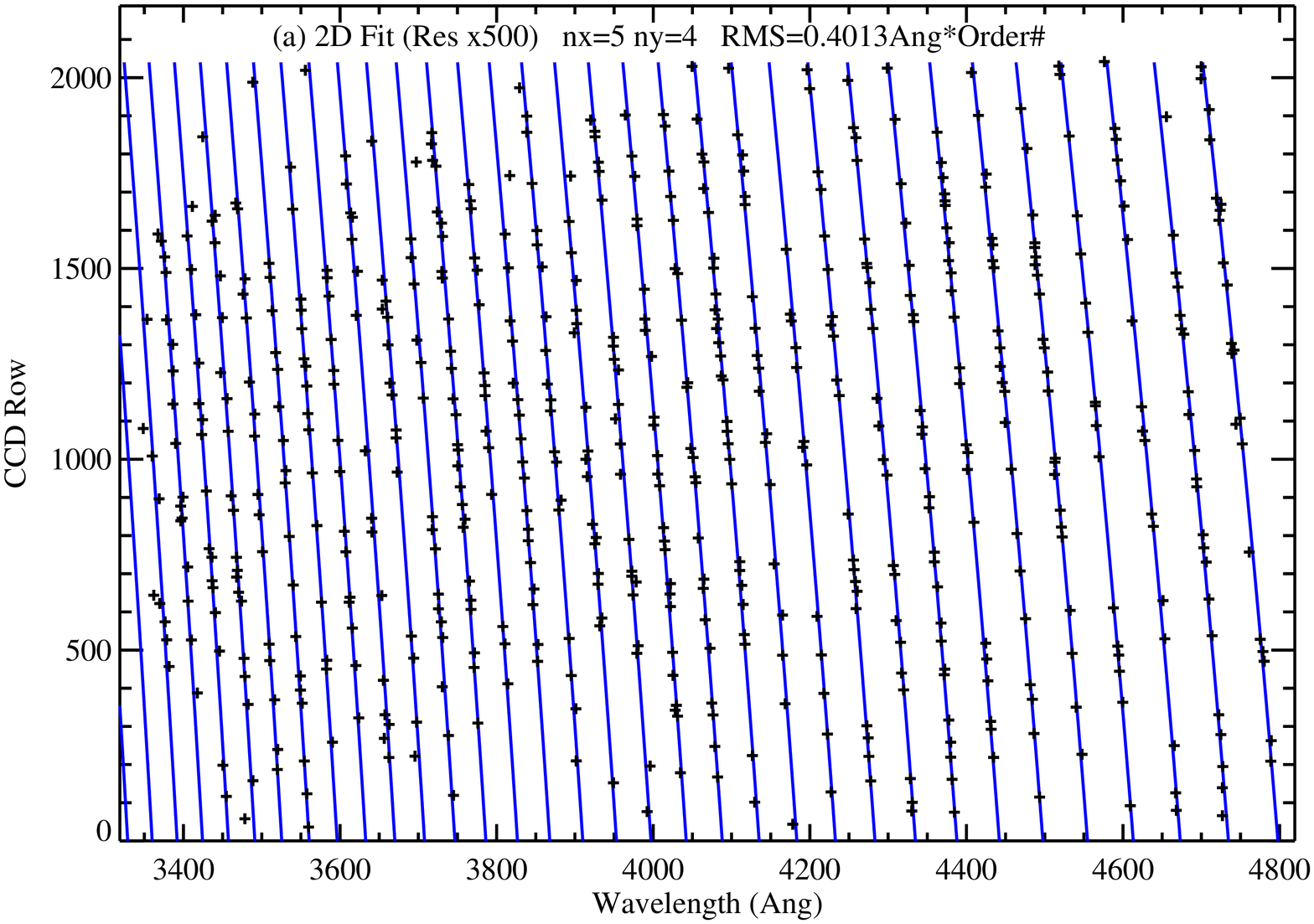}
\includegraphics[height=3.5in,angle=90]{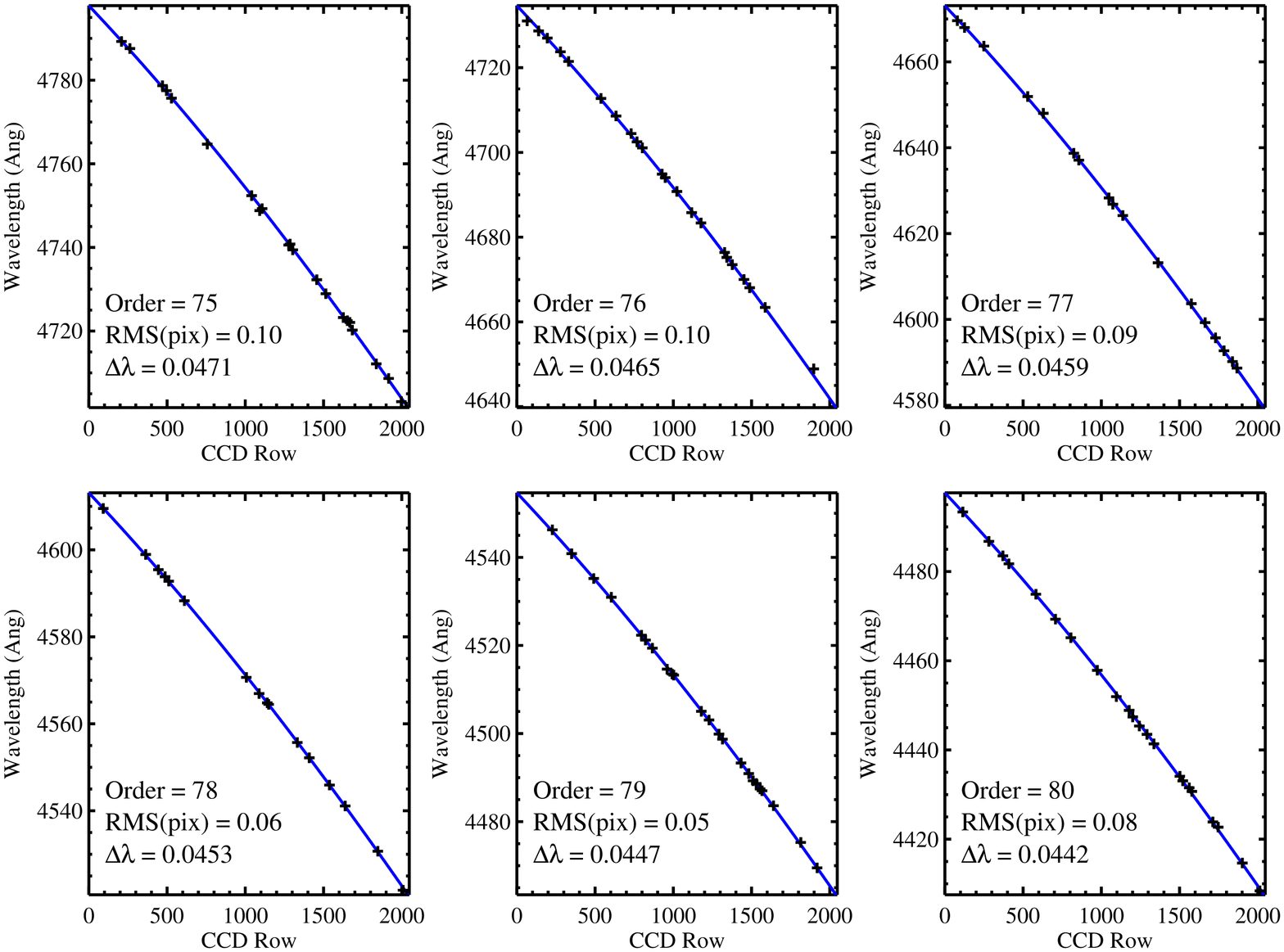}
\caption{(upper) Graphical representation of the 2D wavelength
fit to all of the ThAr lines identified in the ThAr image.
The fit is a 5$\times$4, two-dimensional Legendre polynomial
fit with bases of CCD row and echelle order number.
The blue curves trace the wavelength values, echelle order by
order, as a function of CCD row.  The '+'-signs indicate
the values of individual ThAr lines offset from the 2D solution
by 500$\times$ their residual.
(lower)  Similar to Figure~\ref{fig:wav1D} except the fit here
corresponds to the 2D solution derived for the full image.
}
\label{fig:wav2D}
\end{figure}

The next algorithm performs a 2D fit to all of the good\footnote{Future
updates to the code will include a trimmed ThAr linelist
constructed using the algorithms described in \cite{murphy07}.}
ThAr lines for all of the echelle orders\footnote{Note 
that orders run (mostly) parallel to columns
and with row numbers changing along the order.}. 
The algorithm inputs the pixel (row
number of the line centroid), 
wavelength, and echelle order for each fitted line. 
It then performs a 2D Legendre
polynomial fit $P_{jk}(m,y)$
to the product of the arcline wavelength $\lambda_i$ and its
echelle order $m_i$ as a function of 
(i) the echelle order number and (ii)
the row number $y_i$ (e.g.\ http://iraf.noao.edu/)
\begin{equation}
m \lambda = P_{jk}(m, y) \;\;\;\; .
\end{equation}
With this basis, a well-behaved fit can be derived with 20 
parameters (four spectral and five along the prism dispersion [$j=5,k=4$]);
the typical RMS is $\approx 0.2$(\AA\ order\#).
By performing a 2D solution, one more confidently interpolates
(and occasionally extrapolates)
over regions of the arc image where there is a low density of 
calibrated ThAr lines.
This is particularly important at the reddest wavelengths
where there are many fewer lines for analysis. 

\begin{figure}
\epsscale{1.2}
\includegraphics[height=3.5in,angle=90]{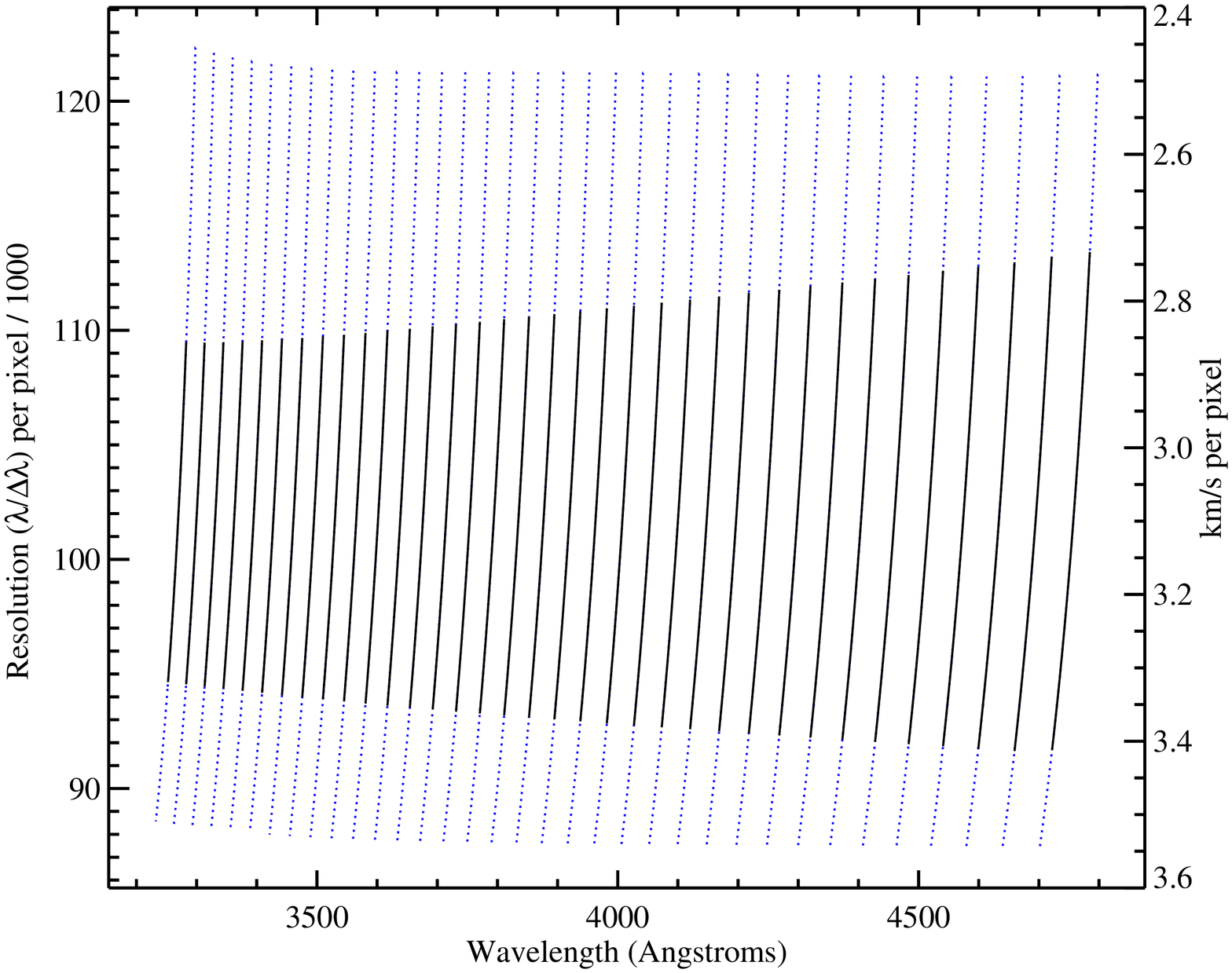}
\includegraphics[height=3.5in,angle=90]{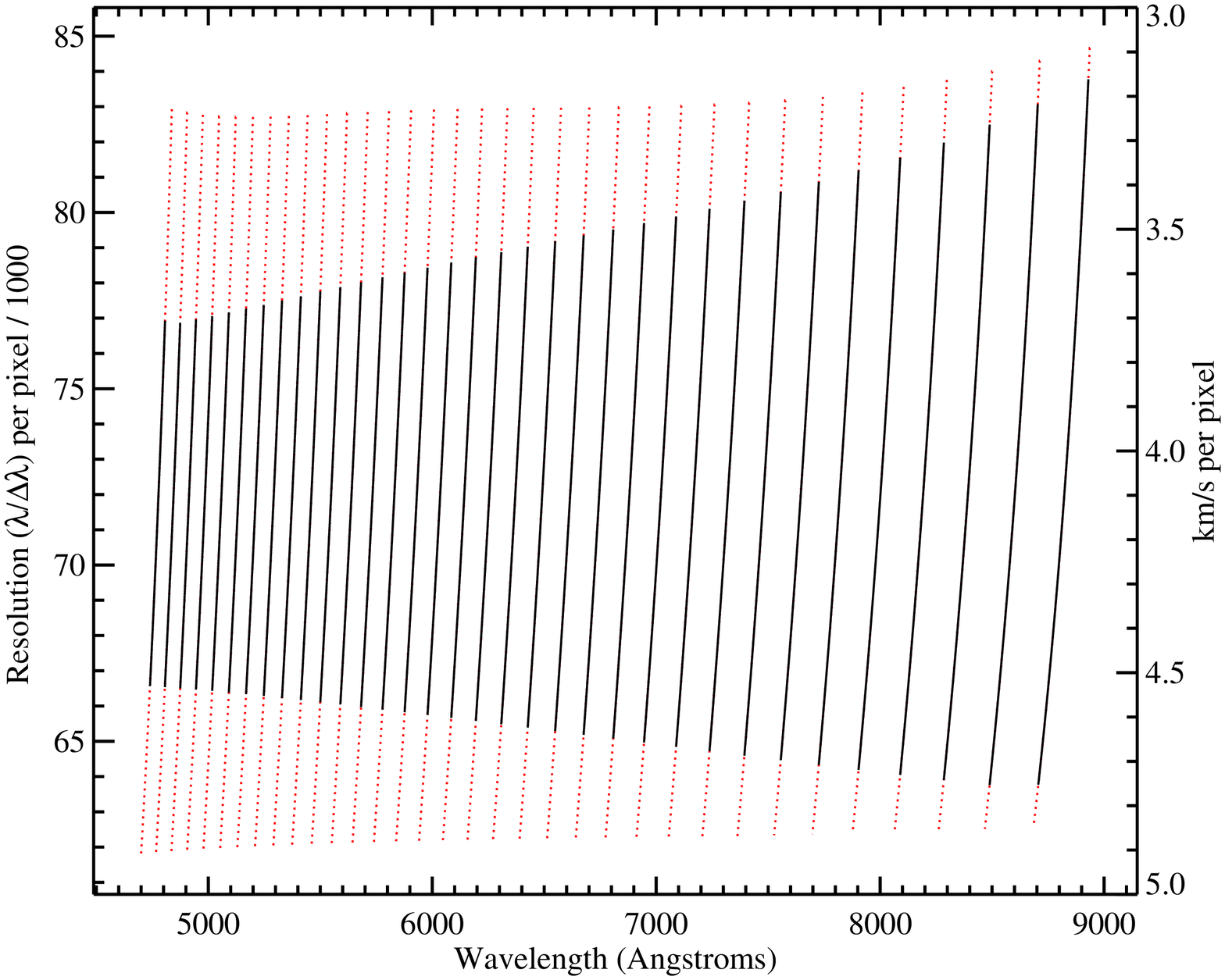}
\caption{Instrumental resolution $\lambda / \Delta \lambda$ for
the native CCD pixels, scaled by 1000, for the red-side (top) and 
blue-side (bottom) cameras of MIKE.  The black regions designate
the 1st order blaze for each echelle order (ranging from number 
71 to 106 on the blue side and 37 to 70 on the red side). }
\label{fig:wavdisp}
\end{figure}

The resulting 2D fit describes the wavelengths along the 
center of each echelle order on the CCD.
Figure~\ref{fig:wav2D}a shows the full 2D solution for one example
image as a function of row number.
Figure~\ref{fig:wav2D}b shows the solution for a subset of 
echelle orders and the measured RMS of the fit.  Again, we
generally achieve residuals of RMS~$< 0.1$\,pix (binned) in each
echelle order for the 20 parameter Legendre polynomial.
Using the 2D solution, we can calculate the
dispersion as a function of wavelength over the entire footprint 
of spectrum. In Figure~\ref{fig:wavdisp}, we present the wavelength dispersion
of the spectrometer as measured for an unbinned spectrum
(native pixels of 15$\mu$m).

\begin{figure*}
\epsscale{0.8}
\includegraphics[height=7.0in,angle=90]{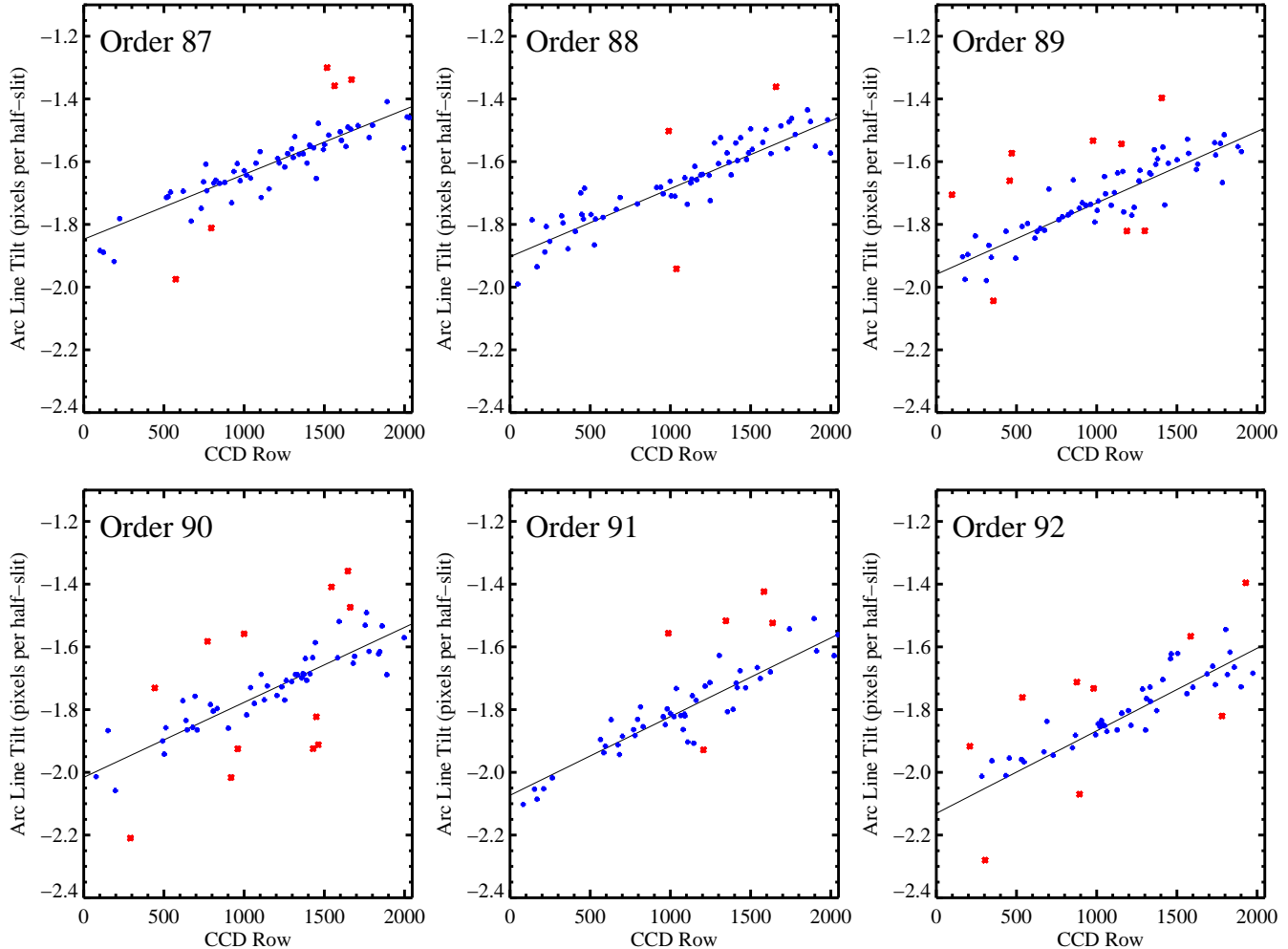}
\caption{Slopes of the ThAr lines measured for several echelle
orders from a representative arc image from the blue-side of MIKE.
The points mark the measured values for each, relatively bright ThAr 
line in the echelle order versus CCD row.  The black line
is a linear fit (with sigma clipping; red points show rejected lines)
to the slopes as a function of CCD row.
}
\label{fig:tilt}
\end{figure*}

\subsection{Wavelength Image}
\label{sec:arcimg}

The algorithms discussed in the previous section 
generate a 2D Legendre polynomial fit describing the 
wavelength value down the center of each echelle order.
We now describe the algorithms which extend the wavelength
solution throughout the full echelle footprint.
The algorithms first identify and trace all high S/N arc
lines in each order across the chip.  In this manner, we measure the tilt
of the lines as a function of wavelength and echelle order.
Specifically, we identify all 5 sigma peaks in the spectrum,
independent of whether the peak corresponds to a well-calibrated 
ThAr line.  We then trace the centroid of each
line from the center of the order to the order edges
using the IDLSPEC2D algorithm {\it trace\_crude}.
We found that each arc line is well described by a straight line
(i.e.\ a slope); the code measures and records the best-fit slope
for each arc line.   
Figure~\ref{fig:tilt} presents the measured slopes 
for the ThAr lines 
as a function of CCD row 
in several echelle orders (blue camera).
In general, one has over 30 lines per echelle order for analysis.  
It is evident that the arc-line slopes vary across each echelle order.  
This is because the quasi--littrow angle ($\gamma$) changes as a function
of position in the focal plane.

Similar to the analysis of the ThAr wavelengths,
we fit a 2D Legendre polynomial
to the slopes of the arc lines across the full CCD.
This 2D solution enables one to 
interpolate through regions of the CCD with a lower density of 
ThAr lines.
Together the wavelength values and arc line tilts
at the center of each order is sufficient 
for assigning a unique wavelength to the center of every pixel
on the CCD that falls in an echelle order footprint.
This is done for all applicable pixels and
a wavelength image is written to the disk for each arc frame.
We do not solve for the case when orders overlap and assume a
unique wavelength value for each pixel in the image.
This wavelength image is calibrated in air and is stored with 
logarithmic values.
By default, the pipeline associates each scientific exposure with
the wavelength image whose arc frame has the smallest difference
in UT time from the start of each exposure.  We do not interpolate
between arc frame exposures, and simply use the nearest exposure in time.

\subsection{Image Shift}
\label{sec:shift}

As described in $\S$~\ref{sec:design}, the footprint of the
MIKE spectrometer is observed to shift throughout the course
of the night, presumably because of thermal changes. 
Therefore, the footprint which is traced by the
flats acquired in the afternoon ($\S$~\ref{sec:tflat})
is generally offset from the one observed during the night.
We measure this shift from the inner 17 orders using
an algorithm similar to the one which measures the slit
profile ($\S$~\ref{sec:slitprof}).  In this case, however,
the algorithm is applied to the arc images.  As with the slit
profile, one must first have a good estimate of the slit
tilt throughout the image.  Therefore, the steps described in
the prior two subsections are first applied to one arc image
(the slit tilt does not vary significantly as the image shifts).

In each order,  the code collapses all pixels
with signal greater than 10 electrons as a function of 
distance from the center of the slit, normalized to
$\pm 1$ by the average order width.  

We then measure the
center of the arc profile by identifying where the flux
dips to $40\%$ of the peak on each side of the profile
and average the two edges.  This center is compared with
zero and gives the shift in pixels of the image parallel
to the rows (the shift along the columns is negligible).
This process is performed on the 17 central orders and the best
linear fit to the shift as a function of order number is found.  
The parameters of this fit are recorded in the main structure 
(in the ARC\_XYOFF tag) and are
used to shift the archived footprint onto the science and
arc images.

%

\section{Object Tracing and Extraction}
\label{sec:object}

\subsection{Tracing}

The final phase in the reduction of two-dimensional CCD images
to 1-dimensional spectra involves the final localization and extraction
of a single object in each MIKE exposure.  The first step involves the tracing
of the object, the definition of the object position with respect to
the spatial position of the echelle orders.  
A first estimate of the local sky background is 
calculated by masking pixels within the object aperture 
(default is the central 3.75$''$ of the 5$''$ slit) and those with
a slit profile value below 30\%.
The remaining steps incorporate an iterative series of 
object profile reconstruction as a function of wavelength and echelle order, 
slight tweaking of the initial object position given the mean 
object profile in each order, and finally, combined object and background
extraction.

Once the echelle orders have been defined by the trace flat images, we determine
the best spatial shifts from the traceflats to the appropriate arcframe
associated with each science image (see \S\ref{sec:shift}).  
Each echelle order in the science frame
is rectified (just for the object tracing) with the shifted order boundaries.
The rectified order is collapsed along columns with median smoothing to reject
unflagged cosmic rays and bad pixels.  In the collapsed vector of each order, 
we locate the most significant peak within the slit
and associate the fractional position 
between the order boundaries as the first guess for the location of 
the object where 0.5 corresponds to the center of the slit-length.
If the object is detected in seven or more echelle
orders, the fractional positions in the undetected orders 
are predicted by a simple linear fit to the detected positions
as a function of order number.  The fractional positions 
for the rectified orders (whether measured
or predicted) are converted back to CCD pixel coordinates on the original 
science image based on the shifted order boundaries.  We measure the object 
peaks in each CCD row in each order by iteratively calculating a flux-weighted
mean of the object counts.  The peak measurement is unbiased for well sampled
profiles, but is subject to CCD artifacts and sharp sky features.  Each
set of object centers, called a trace, 
is fit with a 6th order Legendre polynomial
weighted by estimated errors in the flux weighted centroids (with an additional
floor of 0.01 pixel error added in quadrature).  In the fitting procedure,
centroids which deviate by greater than 10$\sigma$ from the polynomial fit
are masked.  

Echelle orders in which less than 50\% of the measured centroids 
are rejected are considered useful for a subsequent PCA analysis (similar
to the order-tracing analysis described in  \S\ref{sec:tflat}). 
The polynomial coefficients from 1st to 6th order of the well
constrained echelle orders are combined into orthogonal eigenfunctions.
The useful traces are checked to see if their eigenvalues 
(PCA coefficients) agree with simple interpolated predictions of the 
other traces.  The worst outliers with a 1 pixel deviation in the primary 
eigenvalue or a 0.3 pixel deviation in the secondary eigenvalue level are 
rejected.  The remaining eigenvalues are fit as a function of 
echelle order with low-order polynomials to
predict the previously rejected traces, with a quadratic fit to the leading
eigenvalues and a linear fit to the secondary eigenvalues.  The remaining
coefficients are simply replaced with the median value of all the good 
values.

\begin{figure}
\includegraphics[width=3.5in]{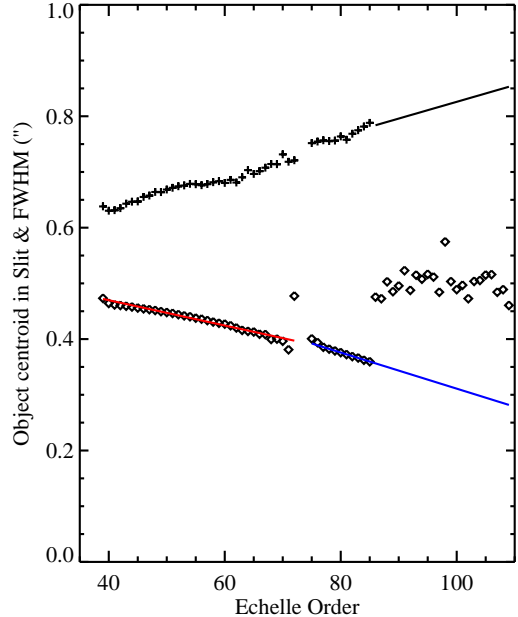}
\caption{Empirical FWHM of the spatial profile and (separately)
the relative object position along slit-length of a high-$z$ quasar
as a function of echelle order for both the blue and red-sides of MIKE. 
The plus signs show the measured FWHM of the spatial profile in arcseconds
assuming a Gaussian profile.  The systematic increase is primarily
due to atmospheric effects.  The diamonds and colored curves
show the measured position of the object along the slit where 0.5
refers to the center of the slit-length.  Note that at echelle orders
beyond 85, the quasar exhibits no significant flux and our algorithm
extrapolates the object centroid. 
}
\label{fig:fwhm}
\end{figure}

\begin{figure}
\includegraphics[width=3.5in]{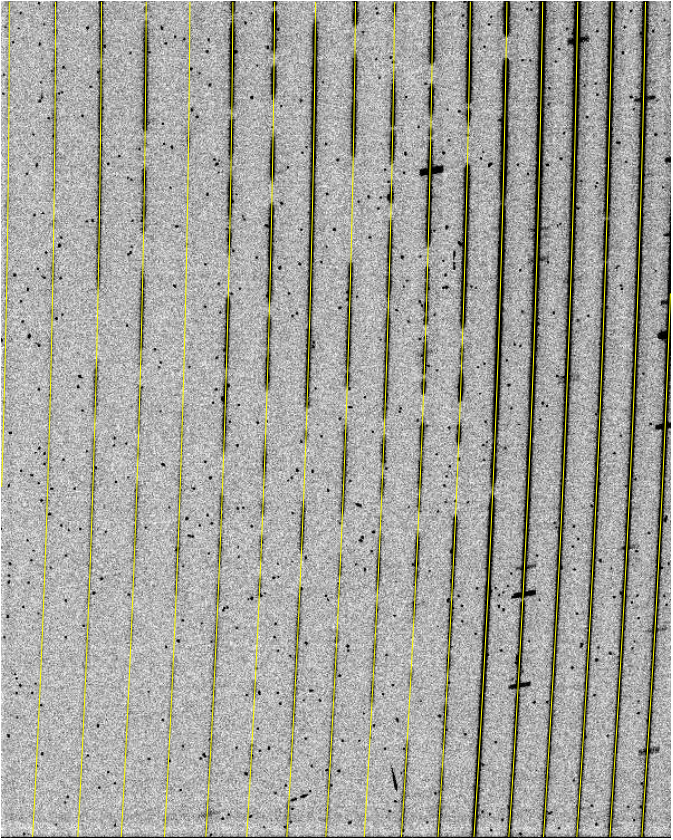}
\caption{Processed 2D image from the red-side of MIKE for a high-$z$ quasar
(black indicates high intensity).  Overplotted on the image is the trace
of the spatial center of the object profile as determined from a PCA
analysis of the full 2D frame.
}
\label{fig:trcobj}
\end{figure}

We calculate the mean trace position (the average spatial position
of each trace over all rows) with the following steps.  First, we remove the
high-order variations (tilt, curvature, etc.) by summing up the contributions
from 1st order to 6th order and subtracting the sum from the measured mean
trace position.  Second, the best quadratic fit to the mean residual trace 
positions are found, and the higher order contributions (orders 1--6) are
added back.  The result is a trace of the object in each echelle order
that relies on the principal components of the well traced orders and
smooth fits to these coefficients.  Finally, we perform a last iteration and
allow the traces which have sufficient signal-to-noise to adopt the 
empirically determined trace centers.  Object traces with low 
signal-to-noise or partial orders retain the interpolated values derived above.
We show how the object trace can
move relative to the center of the echelle order positions 
as the bottom set of open diamonds in Figure~\ref{fig:fwhm}.
The routine invoked for object tracing is called mike\_fntobj.
An example of the final determined positions for an object observed with
MIKE is shown in Figure~\ref{fig:trcobj}.

\subsection{Object Profile Definition}

The next step in object extraction is to estimate the sky
background as a function of wavelength and echelle order number.  
Because this estimate is only used to derive the spatial profile
of the object PSF, accurate fitting of narrow emission lines is not
critical.
Some fraction of the background arises in diffuse scattered light, and
this is estimated by interpolating the pixel counts in the gaps of 
the echelle orders to all of the CCD.  Once the scattered light estimate
has been subtracted, the remaining sky background is fit in each echelle order.
A subset of the pixels in each order are used if they fall at least 1.9$''$ from
the object centroid and have a slit profile value greater than 30\%.
For standard stars, the default is to only use pixels at least 2.25$''$ from the
object, although one is recommended to skip sky subtraction for very
bright targets.  The pixels designated from sky estimation are corrected by their
corresponding slit profile values.  These corrected pixels are then fit 
with a b-spline as a function of wavelength using the wavelength image map 
calculated in \S\ref{sec:arcimg}.  The breakpoints defining the b-spline function
are spaced with a separation of 1.2 times the local wavelength dispersion
per pixel.  After the b-spline fit has converged, sky subtraction is as simple
as evaluating the b-spline function for all pixels in the echelle order, 
correcting by the slit profile (multiplying) and subtracting the product 
from the scattered-light subtracted image.
The routine that performs the initial sky-subtraction on all science images
is called mike\_skysub.

At this stage of the reduction, we finally have a fully characterized 
science image (i.e.\ flat-fielded, traced, 
wavelength calibrated, and background subtracted), 
and can perform optimal extraction in 
the empirical sense \citep[e.g.][]{optimal}
once the object profile is characterized as a function
of wavelength.  With higher signal-to-noise spectra,
we can determine subtle variations in the object profile.
Below is the set of steps we follow to characterize the object profile 
which is then used to optimally extract the 
1-dimensional spectrum in each echelle order. 
The list below is repeated iteratively for a default of three
times, but can be set by the user when invoking the extraction
algorithm mike\_box.

The first step is a standard boxcar extraction in each order, using an aperture
with a default spatial extent of 3.5$''$ centered on the object trace.
The boxcar extraction, along with the associated variance array, is
stored in the final object extraction structure. 
The median signal-to-noise ratio (MSNR) is calculated 
for the boxcar spectrum in each echelle order, and after this point, 
the orders are processed in order of decreasing MSNR. 

The boxcar extraction is median smoothed and compared to the original spectrum
to identify cosmic rays and other deviant spectral pixels.  The non-deviant 
pixels are fit with a b-spline function with a breakpoint spacing that is twice
the local pixel dispersion (i.e.\ every 2 spectral pixels).
In this step, we are fitting for the spatial profile, so we normalize each 
CCD pixel in the 2D image by the boxcar extracted counts at the appropriate 
wavelength (the bspline function representing the boxcar extracted flux is
evaluated at every pixel in the order footprint).  
The pixels in the resulting normalized image have, by definition, a total 
spatial cross-section of unity in the boxcar region.
This definition is one method to tie the optimally extracted fluxes to the same 
system as the boxcar extracted values. 

The object profile is fit with one of three methods based solely on the
MSNR determined for that echelle order.  If the MSNR is less than 2.5, 
then a uniform Gaussian profile is adopted with a fixed width.  If at least 2 
higher MSNR orders have been extracted (earlier in the processing), then 
the width of the Gaussian profile is extrapolated from the processed orders.
If an estimation of a width based on other profiles is not available, then a
fixed width of 0.7$''$ is assumed (comparable to the median seeing at
Las Campanas).  If the MSNR is greater than 2.5, then a
Gaussian is not assumed, and a full spatial profile is fit.  If the square
of MSNR is greater than 500, then the spatial profile is allowed to vary as
a quadratic function of wavelength along the order.  
Otherwise, a uniform profile is fit to all pixels in the order.  
Even in the high MSNR case, a uniform profile
is fit to the order, but the full variable profile is used in object 
extraction.  The full-width, half-maximum (FWHM) is calculated empirically
from the uniform profile and is used as a prior for subsequent lower MSNR 
orders.  In Figure~\ref{fig:fwhm}, the calculated and extrapolated 
estimates of the spatial FWHM are shown as a function of echelle order.

\begin{figure}
\includegraphics[width=3.5in]{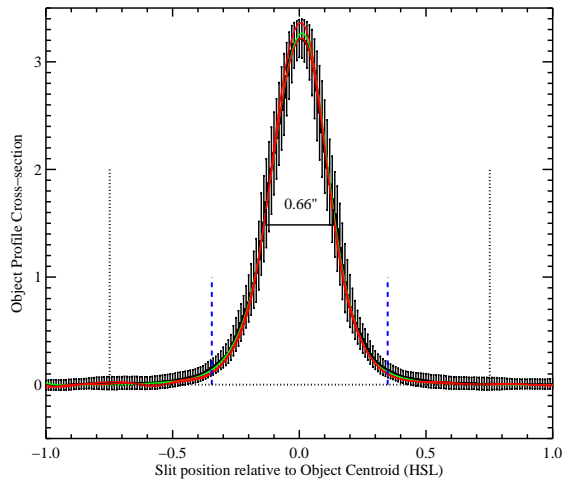}
\caption{An example of the object profile for a single echelle order.  
The ticks
are the boxcar normalized data, showing the 50\% ranges as a function of
half-slit length.  The green line (middle line) shoes the mean
object profile in the order, while the red lines show the profile at the very
top and bottom of the order.
The dotted-black vertical lines show the boxcar aperture, and the 
dashed-blue vertical lines
show the positions where the profile extrapolation begins.  
The FWHM in arcseconds (here 0.66$''$) 
is determined empirically by locating the points of half-maximum in the 
mean order profile.
}
\label{fig:profile}
\end{figure}

The biggest difficulty with an empirical determination of the object profile 
are the wings of the profile.  These need to be well behaved to perform
accurate extractions.  In this implementation, we specify a separation from the
object centroid, past which we extrapolate an exponentially decaying wing.
The separation increases as a function of MSNR, such that higher MSNR orders
require less extrapolation.  To extrapolate, we find the best linear fit
to the logarithmic profile as a function of spatial distance 
from the object centroid.
We perform the extrapolations on left and right sides of the profiles 
separately and in each case we fit to the shape of the profile from the 
half-maximum to the cutoff distance determined by the MSNR above.
This procedure is very robust in practice and delivers realistic wings for
the object profiles.  The slope of the extrapolation is forced to be less
than $-1.0$, because a higher slope would not have finite flux if 
extended to infinity.  In Figure~\ref{fig:profile}, we show an example of the
object profile determined in a single MIKE echelle order.  The ticks denote the
extent of the scatter in the normalized image counts, while the solid lines
represent the spatial object profile fit as described by a b-spline.

\subsection{Extraction}

Just before the actual extraction occurs, we perform
a check and correction to the object tracing.
Now that we have a well determined empirical profile for the order, we test
the correlation of the profile at the current position compared to shifting
the profile one pixel left or right.  This check is done at each row in the
echelle order, and we apply a smooth fit of the
deviation away from zero offset.
The trace correction is added to the object centroid array and
is applied in the next iteration.

\begin{figure}
\epsscale{0.9}
\includegraphics[width=3.5in]{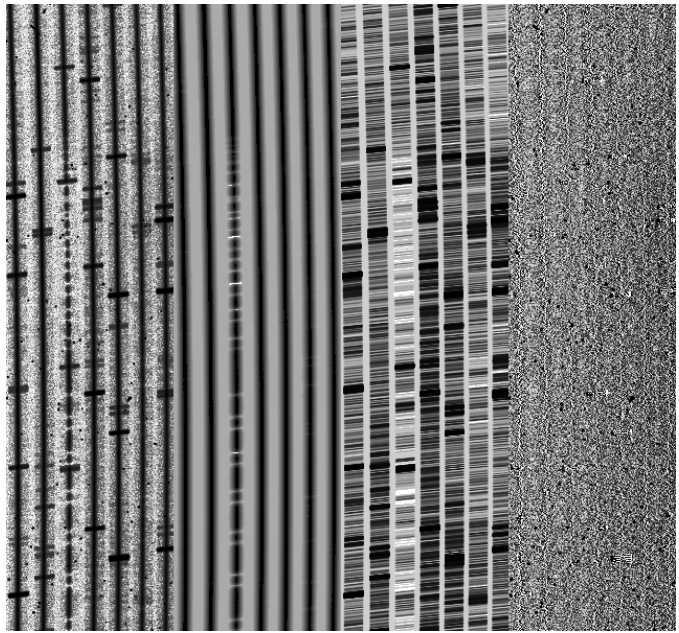}
\caption{Four panels showing the extraction results on the red side of a
science observation of PKS2000$-$330 \citep[see][]{ppo+10}.  The left most panel
is the flattened, scattered light subtracted science frame.  The 2nd and 3rd
panels show the two-dimensional models of the object and sky spectra, 
respectively.  The right most frames show the residuals about the best fit 
models, which are close to the Poisson limit in all pixels except the brightest
sky features. }
\label{fig:extract}
\end{figure}

Extraction is performed by fitting in the least squares sense 
to the full echelle order footprint with a simple linear model:

\begin{equation}
Model(x,\lambda)= OS(\lambda) * OP(x,\lambda)+ SS(\lambda) * SP(x,\lambda),
\end{equation}
where x is in units of HSL, and $\lambda$ is wavelengths.  OS and SS are the
object spectrum and sky spectrum, respectively, while OP and SP are the 
previously determined object profile and slit profile, respectively.
The extraction is a simultaneous b-spline fit of both the object spectrum and
the sky spectrum.  The breakpoints for each spline are separated by exactly
the local wavelength dispersion. For a standard MIKE echelle order
which is 2048 rows high (spectral binning of 2), there will be approximately
4096 free parameters in the simultaneous b-spline fitting.  
We then search for pixels deviating by greater than 10-sigma, and mask those
pixels as well as the 12 closest neighbors
under the assumption that these are cosmic rays.
During the final iteration, one last b-spline fit is performed with the final
masking, and both the object and sky splines are sampled with uniform 
velocity dispersion on a fixed wavelength grid to facilitate 
combining multiple science exposures (Figure~\ref{fig:extract}).  One can output
the model fits of the two-dimensional sky image,
 object image and object profile. 

\section{Endgame}
\label{sec:endgame}

In this section, we discuss the procedures which combine and 
flux calibrate the extracted spectra (in separate echelle orders).
The end product is a continuous, 1D spectrum calibrated to 
energy units with a relative accuracy of better than $10\%$ over
100\AA\ spectral regions.
In practice, this involves combining the 
multiple exposures of a single target, fluxing, and then 
coadding the individual echelle orders.  We present separate
discussions of fluxing and coaddition.

\subsection{Fluxing}

Spectra acquired with echelle spectrometers are notoriously
difficult to calibrate in absolute flux units.
This can be attributed to a number of factors:  narrow slits,
scattered light, vignetting.
It is generally difficult to achieve even
an accurate relative fluxing of the data \citep[e.g.][]{dh,elp+99,suzuki}.
Furthermore, little attention is given to this problem
because the analysis of echelle data usually involves
normalizing the object's continuum flux level prior
to measuring the optical depth and/or equivalent width of spectral
features.  In this respect, fluxing is mainly a convenience but
not a necessity.  Nevertheless, the MIKE instrument has several
characteristics which make it better suited to accurately flux (see the
Introduction).

\begin{figure}
\includegraphics[height=3.5in,angle=90]{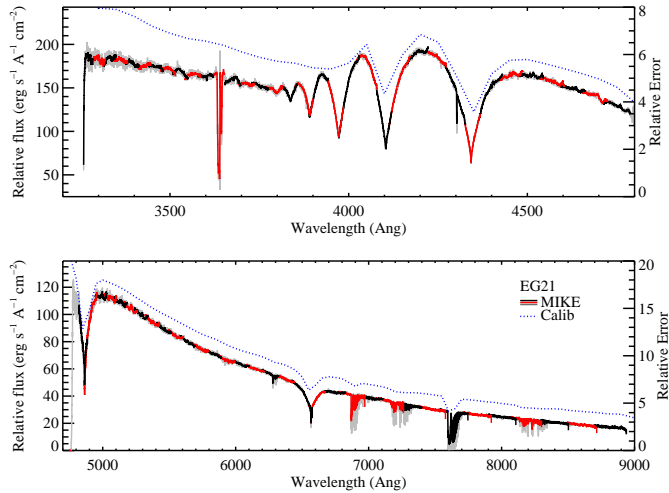}
\caption{MIKE echelle spectrum of the spectrophotemetric
standard EG21, calibrated with a sensitivity function derived
from analysis of NGC~7293.  The top panel shows the data
obtained with the blue camera and the bottom panel shows data
acquired with the red camera.  
The alternating red and black data
indicate independent echelle orders (flux and error), 
smoothed by 15 pixels for presentation purposes.  The gray curve 
traces the flux and error of the final, combined spectrum.
One notes that there is excellent agreement in the regions of
order overlap.  Furthermore, the Balmer lines, which span 
several echelle orders, have physically appropriate shape
and relative equivalent widths.
The dotted blue curve shows the relative flux of EG~21 as
calibrated by \cite{hamuy94}.  This curve
has been scaled to match the MIKE spectrum at 4600\AA\
and 6000\AA\ and offset by 20 and 10 units respectively for the
blue and red data.  One notes the relative flux is accurate to
well within 10$\%$ for all wavelengths longer than 4000\AA.
At bluer wavelengths, the MIKE spectrum underestimates the flux
by an increasing amount due to the uncorrected slit loss and 
additional atmospheric extinction as EG~21 was observed at
higher airmass than NGC~7293 (AM=1.36 vs.\ 1.1).
}
\label{fig:eg21}
\end{figure}

We implement algorithms which follow standard procedure to 
generate a sensitivity function from observations of a spectrophotometric
standard.  This sensitivity function converts measured electrons per
second per \AA\ to flux units (erg s$^{-1}$ \AA$^{-1}$ $\cm{-2}$).
It is determined by comparing the electron flux of the spectrometric
standard with its absolute, previously calibrated flux.
No corrections are made for slit losses or atmospheric extinction.
These efforts are relatively shallow functions of wavelength and should
not substantially affect the relative flux on scales smaller than 100\AA.
Furthermore,
if one observes the science objects and standard under similar seeing
conditions and at similar airmass, then these losses will be 
partially corrected.

The only significant challenge that we faced is that no spectrophotometric
standards have been observed at echelle resolution.  
As such, we found it impossible to derive an accurate sensitivity 
function in spectral regions corresponding to
absorption features in the standard star (e.g.\ the Balmer series).
In low dispersion spectrometers, one can clip these features
by fitting a low order polynomial to the sensitivity function. 
In echelle data, however, a strong Balmer
line can encompass one or more echelle orders.   For this reason,
one should observe calibration standards with weak
absorption lines.  We also designed a graphical user interface
which enables the user to identify and mask out absorption features
when generating the sensitivity function.  This is generally 
successful, although systematic errors are significant when the
features coincide with the peak or edge of an echelle order.

\begin{figure}
\epsscale{0.8}
\includegraphics[height=3.5in,angle=90]{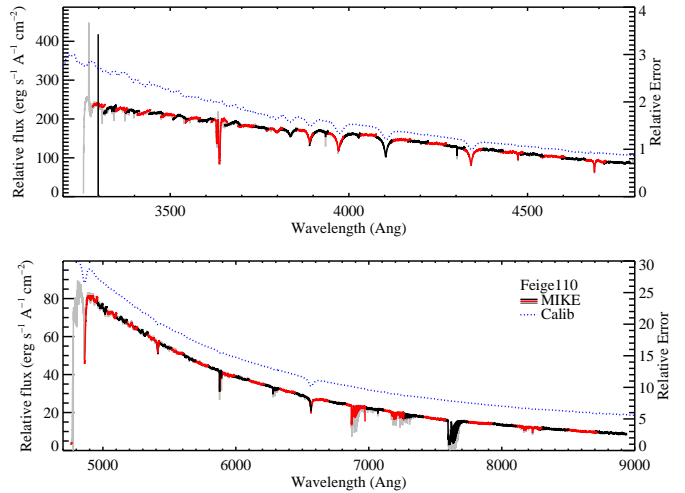}
\caption{Similar to Figure~\ref{fig:eg21} but for Feige~110.
These results are more typical of standard performance.  There is mild
echelle order 'mismatch', especially in the first $\approx 10$
orders of each camera.  The relative flux is very accurate for
data redward of 4000\AA, but is underestimated at
lower wavelengths.  This is the result of additional slit loss
and higher airmass than the calibration standard.
}
\label{fig:feige110}
\end{figure}

In Figures~\ref{fig:eg21} and \ref{fig:feige110}
we present the fluxed spectra of two standard stars observed
on 02 September 2004 with a 1$''$ slit under photometric skies.
We calibrated these stars (as described above) by generating
a sensitivity function from the spectrophotometric standard
NGC~7293 observed at 1.11 airmasses.
The spectrum of EG21 (Figure~\ref{fig:eg21}) represents
a nearly optimal example.  The flux in overlapping regions of
echelle orders is well matched, the strong Balmer lines have 
appropriate shape, and the relative flux matches the 
calibrated values \citep{hamuy94} to better than 5$\%$ for
$\lambda > 3800$\AA.  At bluer wavelengths, the flux is underestimated
because of additional slit losses and atmospheric 
extinction;  the star was observed at a higher 1.36 airmasses.
The results for Feige~110 (Figure~\ref{fig:feige110}) are more
representative.  There is a modest mismatch
in the flux of overlapping echelle orders, especially the
first $\approx 10$ orders in each camera.  The mismatch is less
than 10$\%$, however, and the combined spectrum (grey) follows
the `true' calibration.  The relative fluxing is better than 10$\%$
redward of 4000\AA, and the spectrum underestimates the
flux at bluer wavelengths for the same reasons as EG~21.
We emphasize, however, that the relative flux is excellent over
spectral regions of less than 100\AA.

\begin{figure}
\includegraphics[height=3.5in,angle=90]{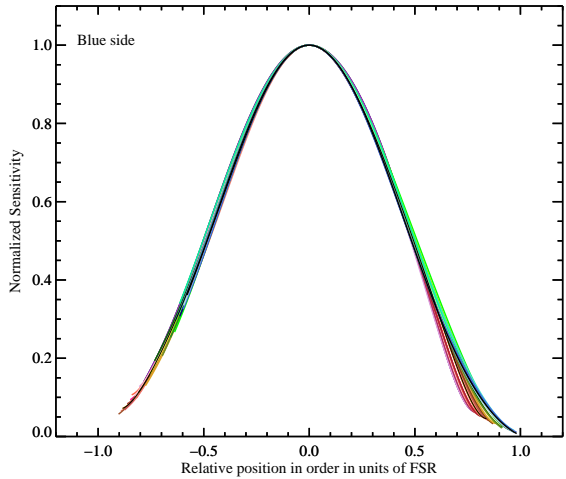}
\caption{Normalized blaze function for the blue side of MIKE.
}
\label{fig:blaze}
\end{figure}

\begin{figure}
\includegraphics[height=3.5in,angle=90]{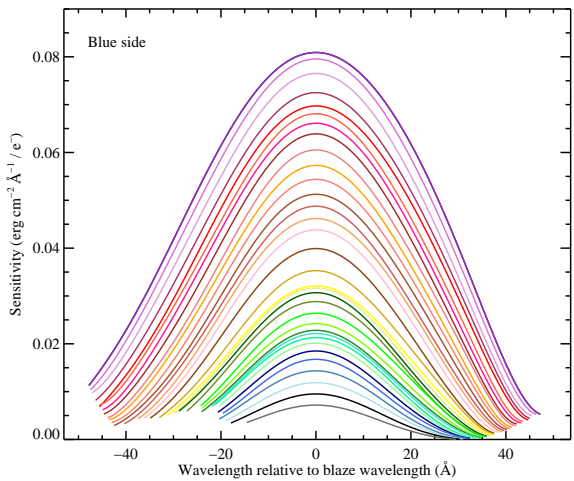}
\caption{The sensitivity function of each order on the blue side of
    MIKE is shown as a function of wavelength relative to the central
    wavelength of the order.  If the sensitivity function of each
    order is normalized to 1 at the central wavelength, and the
    wavelength range is normalized to the free spectral range (i.e. if
    the wavelength range is scaled to ${\pm 1}$), then the blaze
    functions shown in Figure \ref{fig:blaze} are obtained.
    }
\label{fig:sensitivity}
\end{figure}

It is instructive to analyze the sensitivity function to 
investigate the blaze function and instrumental response
of MIKE.  Figure~\ref{fig:blaze} describes the blaze function,
normalized to unit flux at peak, for the blue camera as
derived from the sensitivity function.  The blaze function
profile is nearly constant with echelle order when plotted
as a function of the free spectral range (FSR).  The shape
is well modeled by a 9th order Legendre polynomial.
which is the default function form for fitting the sensitivity
function, which is shown in Figure~\ref{fig:sensitivity}.

\subsection{Coaddition}

There are two important aspects to producing a final 1D
spectrum from a series of science exposures:
the coaddition of multiple exposures and the combination
of overlapping echelle orders.
The specific ordering of our algorithms are:
(1) coadd multiple exposures of each echelle order;
(2) flux the individual echelle orders;
and (3) combine the echelle orders to produce a final, continuous
1D spectrum.
We prefer this ordering because it maximizes the signal-to-noise 
for combining echelle orders, whose blaze edges may have relatively
low signal.
Furthermore, the instrument is sufficiently
stable that one can successfully combine individual echelle
orders from multiple nights in an observing run without fluxing first.
The fluxing algorithms were discussed above;
we now detail the coadding procedures.
We remind the reader that the spectra have been extracted onto
a fixed wavelength grid.  No rebinning is required and registration
of the spectra is trivial.

The combination of multiple science exposures of the same object
is performed separately on each echelle order, in sequence of
red to blue.  The orders are averaged together, weighting by the square
of the median S/N ratio after first flagging outliers 
(generally due to low-level cosmic rays).
To carry out this calculation, one requires two quantities:
(i) the median S/N ratio for weighting the data; and
(ii) a scale factor which normalizes the spectra to a common
intensity.  The latter is necessary for properly averaging 
data when clipping outliers.
These quantities are calculated relative to a designated `template' 
exposure, preferably the frame with largest signal.

\begin{figure}
\includegraphics[height=3.5in,angle=90]{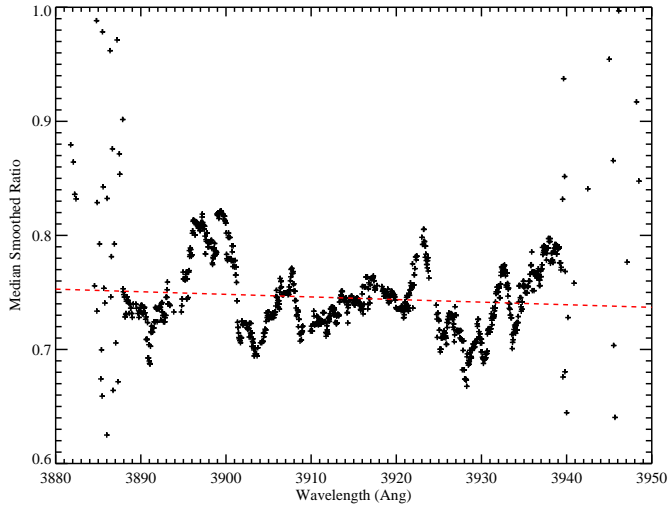}
\caption{Ratio of two extracted spectra, median smoothed by
100 pixels,  from echelle order 91 of the blue-side of MIKE.
Values are shown only at the position where individual pixels 
have S/N~$>2$. The dashed red line shows a linear fit to the
median smoothed ratio values.
This is used to scale the spectra prior to coaddition.
}
\label{fig:hsnpix}
\end{figure}

All pixels with S/N~$>2$ (designated HSNPIX here) are identified
in the template exposure.  If there are over 100 HSNPIX, then
the median S/N is measured from these pixels.  Otherwise, the
code adopts weighting factors based on the relative exposure times.
The algorithm also calculates a scaling factor by comparing the
fluxes of the HSNPIX for each exposure relative to the template.
Figure~\ref{fig:hsnpix} presents an example of the relative flux
for echelle order 91 of the blue camera for a pair of 2400s exposures
taken in sequence.  We show only the HSNPIX and have median smoothed
by 100 pixels.  The median S/N per pixel is 6 for the template
exposure.   The algorithm scales the data differently depending
on the number of HSNPIX.
If there are over 300 HSNPIX, a line is fitted to the relative
fluxes (median smoothed by 100 pixels)
as a function of wavelength and applied to the full spectrum  
(Figure~\ref{fig:hsnpix}).
If there are 100 to 299 HSNPIX, then the scale factor is the median of 
the relative fluxes.  For fewer than 100 HSNPIX, we adopt the relative
exposure times as the scale factors.

The median S/N and median scale factors are recorded for each
echelle order, in turn.  After all of the echelle orders are
analyzed, the algorithm re-evaluates the two quantities for those
orders with fewer than 100 HSNPIX.  The code takes the median values
of the two quantities from the nearest 5 orders that have (i)
greater than 100 HSNPIX and (ii) lie within 10 orders, if these
conditions are met.  This is implemented primarily on
orders where a very large absorption feature (e.g.\ a damped \lya\ line)
has significantly depressed the S/N of a single echelle order.

After scaling the data, the algorithm compares the scatter
between multiple echelle orders with the scatter predicted from
the variance array.  As described in $\S$~\ref{sec:proc},
the variance array includes
estimates of the uncertainties associated with counting statistics and read noise.
In general, we find that the scatter predicted from the variance array is 
10 to 20$\%$ lower than the observed scatter in multiple exposures.
This could be explained by a modest overestimate of the CCD gain,
but the algorithms which derive the gain are proper.
We suspect that the additional scatter
is due to systematic errors introduced by flat-fielding, 
sky subtraction, and wavelength calibration.  Our solution is to
scale the variance arrays of all the exposures of a given echelle order
by a single factor.  In this manner,  we demand that the 
observed scatter is consistent
with that predicted from the variance arrays assuming $\chi^2$
statistics.

If there are three or more exposures then the code flags and excludes
any pixels that deviate by more than 5$\sigma$ from the median value.
Clipping is rare for optimally extracted data (or short exposures of bright
targets) and it is not necessary to repeat this step.
Finally, the data are averaged with weights proportional to 
the median S/N squared.
The resulting data product is a series of combined, individual echelle
orders.  These data are fluxed with the sensitivity function derived
from a spectrophotometric standard, as described in the previous section.

The final step is to combine the edges of overlapping echelle orders.
Again, no rebinning is necessary and registration is trivial because
the data were extracted onto a fixed wavelength grid.
In the previous sub-section, we noted that the relative flux of the
overlapping spectral regions generally agree to within 10$\%$
(Figures~\ref{fig:eg21},\ref{fig:feige110}).  Therefore, we simply 
average the data, weighting by the variance of each pixel.
Aside from a small spectral region, one of the echelle orders will
dominate the signal in the overlap region.

\section{Summary}

We have described the software package MIKERedux designed and
developed to reduce the echelle images produced by the MIKE
spectrometer.  This code is publicly available and its primary
algorithms have been applied to data reduction for
other modern spectrometers.   While development of this package has
halted, one of us (JXP) plans to release a Python-based code following
many of the procedures described here.

\acknowledgments
The strategies and implementations discussed here are the result 
of many fruitful conversations with our colleagues.  In particular, 
JXP acknowledges Dan Kelson for inspiring some of the algorithms
described here during the time they overlapped at Carnegie, and RAB
thanks Steve Shectman for the countless days spent discussing all
aspects of (and building) MIKE.  We further thank R. Cooke for his 
comments and criticisms on a draft of the manuscript.

\clearpage

\end{document}